\documentclass[11pt]{article}

\usepackage{epsfig}
\usepackage{times}
\usepackage{epstopdf}
\usepackage{amsmath}
\usepackage{amssymb}
\usepackage[left=1.75cm,right=1.75cm,top=3.0cm,bottom=3.0cm]{geometry}
\usepackage{setspace}

\begin{document}

\bigskip\bigskip
\centerline {\Large \bf {Momentum in General Relativity: Local versus Quasilocal Conservation Laws}}
\bigskip\bigskip
\centerline{\large Richard J. Epp$^{a}$, Paul L. McGrath$^{a}$ and Robert B. Mann$^{a,b}$}
\bigskip\bigskip
\centerline{${}^a$ \em Department of Physics and Astronomy, University of Waterloo, Waterloo, Ontario N2L 3G1, Canada}
\vspace{0.2 cm}
\centerline{${}^b$ \em Perimeter Institute for Theoretical Physics, Waterloo, Ontario N2L 2Y5, Canada}
\vspace{0.3cm}
\centerline{\em  rjepp@uwaterloo.ca, pmcgrath@uwaterloo.ca, rbmann@sciborg.uwaterloo.ca}
\bigskip\bigskip

\begin{abstract}
We construct a general relativistic conservation law for linear and angular momentum for matter and gravitational fields in a finite volume of space that {\it does not rely on any spacetime symmetries}. This work builds on our previous construction of a general relativistic energy conservation law with the same features~\cite{EMM2012}. Our approach uses the Brown and York~\cite{BY1993} quasilocal stress-energy-momentum tensor for matter and gravitational fields, plus the concept of a rigid quasilocal frame (RQF) introduced in references~\cite{EMM2009,EMM2011}. The RQF approach allows us to construct, in a generic spacetime, frames of reference whose {\it boundaries} are rigid (their shape and size do not change with time), and that have precisely the same six arbitrary time-dependent degrees of freedom as the accelerating and tumbling rigid frames we are familiar with in Newtonian mechanics. These RQFs, in turn, give rise to a completely general conservation law for the six components of momentum (three linear and three angular) of a finite system of matter and gravitational fields. We compare in detail this quasilocal RQF approach to constructing conservation laws with the usual local one based on spacetime symmetries, and discuss the shortcomings of the latter. These RQF conservation laws lead to a deeper understanding of physics in the form of simple, exact, operational definitions of gravitational energy and momentum fluxes, which in turn reveal, for the first time, the exact, detailed mechanisms of gravitational energy and momentum transfer taking place in a wide variety of physical phenomena, including a simple falling apple. As a concrete example, we derive a general relativistic version of Archimedes' law that we apply to understand electrostatic weight and buoyant force in the context of a Reissner-Nordstr\"{o}m black hole.
\end{abstract}

\section{Introduction and Summary} \label{Introduction}

With the advent of general relativity, the notions of the energy and momentum (linear and angular) of a system became elusive. There are two reasons for this. Firstly, when the spacetime geometry is treated as a dynamical field, energy and momentum are no longer local concepts, i.e., there is no such thing as an energy or momentum per unit volume, which when integrated over a finite volume yields the total energy or momentum inside that volume. As we will discuss below, this non-localizable nature applies to both gravitational {\it and} matter fields. In other words, the local stress-energy-momentum tensor of matter, $T^{ab}$, is not, in any fundamental way, related to the matter energy or momentum of a finite system. Secondly, energy and momentum are frame-dependent constructs. In Newtonian space-time we have the concept of an inertial reference frame that allows us to define the energy and momentum of a point particle or an extended system {\it relative} to such a frame. We can even use a frame that is accelerating or rotating, provided we properly account for non-inertial effects. Thus, the key concept is not that of an {\it inertial} frame, but that of a {\it rigid} frame, that is, one in which the distances between all nearest-neighboring pairs of observers comprising the frame are constant in time. In Newtonian space-time, such rigid frames have precisely six arbitrary time-dependent degrees of freedom: three for linear velocity and three for angular velocity. Such rigid frames also exist in special relativity, but their degrees of freedom are restricted to either arbitrary time-dependent linear velocity with {\it no rotation} (``plane motions"), or motion of the frame generated by a Killing vector of the flat spacetime (``group motions"), e.g., constant rotation with one point fixed~\cite{SalzmanTaub1954}. In general relativity the situation is worse: in general, {\it no} such rigid frames exist. This presents a serious obstacle to constructing physically sensible and useful definitions of energy and momentum in the context of general relativity, and their related conservation laws.\footnote{In general relativity there exist notions of energy and momentum for an isolated system in a spacetime that admits asymptotic symmetries at infinity, but these are a throwback to the pre-general relativistic practice of relying on spacetime symmetries to construct conservation laws. This paper represents a break from that tradition.}

The solution to the first problem (failure of the notion of local energy or momentum) involves shifting from a local to a {\it quasilocal} way of thinking. To see this, let $I_{\rm mat}[g,\varphi]$ denote an action functional for a set of dynamical matter fields, $\varphi$, in a spacetime $\mathcal{M}$ with non-dynamical (fixed) background metric $g$. Since the metric is not treated as a dynamical field, $I_{\rm mat}[g,\varphi]$ is the {\it total} action functional. If, as is usually done, we define the total local stress-energy-momentum tensor of the system as the functional derivative of the total action functional with respect to the metric, we have
\begin{equation}\label{T_matter}
2\delta_g \, I_{\rm mat}[g,\varphi]=\int_\mathcal{M}\epsilon_\mathcal{M} \,T^{ab}\,\delta g_{ab}
\end{equation}
(where $\epsilon_\mathcal{M}$ is the spacetime volume form), and so the total local stress-energy-momentum tensor of the system is just the matter stress-energy-momentum tensor, $T^{ab}$. Insofar as a conservation law constructed from $T^{ab}$ will be homogeneous in $T^{ab}$, it will be essentially blind to any interesting gravitational physics. However, in general relativity the metric is treated as a dynamical field, and we must add to $I_{\rm mat}[g,\varphi]$ the action functional of the gravitational field. Using the usual first order action functional for gravity we then find
\begin{equation}\label{T_matter_and_gravity}
2\delta_g \, I_{\rm mat+grav}[g,\varphi]=\int_\mathcal{M}\epsilon_\mathcal{M} \, \left(T^{ab}-\frac{1}{\kappa}G^{ab}\right)\,\delta g_{ab}+\int_{\cal B}\epsilon_{\cal B} \, \left(-\frac{1}{\kappa}\Pi^{ab}\right)\,\delta \gamma_{ab}
\end{equation}
(where $\kappa=8\pi G/c^4$ and $G^{ab}$ is the Einstein tensor), which tells us that the total local stress-energy-momentum tensor of the system is the sum, $T^{ab}-\frac{1}{\kappa}G^{ab}$, which is just zero by the Einstein equation, i.e., there is no nontrivial local notion of total stress-energy-momentum in general relativity. Note that this statement applies to both gravitational {\it and} matter fields, not just gravitational. This argument is not new. The idea that $-\frac{1}{\kappa}G^{ab}$ is the local stress-energy-momentum tensor of the gravitational field was independently put forward by both Lorentz and Levi-Civita, but was rejected by Einstein on various physical grounds, e.g., gravitational waves in vacuum could then not transport energy, and Einstein and others continued to use a pseudotensor to represent local gravitational stress-energy-momentum~\cite{Cattani_and_De_Maria_1993}. We know today that what saves us is the boundary term. On the right-hand side of equation~(\ref{T_matter_and_gravity}), $\Pi^{ab}$ is the gravitational momentum conjugate to the three-metric $\gamma_{ab}$ induced on the boundary, $\cal B$. In the spirit of identifying the stress-energy-momentum tensor as the functional derivative of the action with respect to the metric, one identifies $-\frac{1}{\kappa}\Pi^{ab}$ as the {\it quasi}local total stress-energy-momentum tensor in general relativity. It is defined only on the boundary (energy and momentum per unit {\it area}), and includes contributions from both the matter and gravitational fields. $T^{ab}$ is no longer involved. This is the essence of what Brown and York did in 1993~\cite{BY1993}.

The solution to the second problem (failure of the general existence of rigid frames) also involves shifting from a local to a quasilocal way of thinking. Local thinking suggests that a finite reference frame is a three-parameter family of observers who fill a finite volume of space, and whose worldlines sweep out a four-dimensional worldtube volume of spacetime with three-boundary $\mathcal{B}$. To say that this frame is {\it rigid} is to say that the orthogonal distances between all nearest neighboring pairs of observers' worldlines remain constant in time. This represents six differential constrains (the vanishing of the symmetric spatial strain rate tensor) on three functions (the three independent components of the observers' four-velocity), which is an over-determined system that in general has no solutions. However, in a Newtonian space-time, where the orthogonal distance between worldlines is just the instantaneous spatial distance, the rigidity condition {\it does} have solutions, namely, the six-parameter family of (not necessarily inertial) rigid frames we are familiar with in Newtonian mechanics. In special relativity, on the other hand, the orthogonal distance is measured with respect to the flat Lorentzian metric, and problems arise due to the relativity of simultaneity, problems which are exacerbated in the curved Lorentzian metric of general relativity. As we discovered in references~\cite{EMM2009,EMM2011}, the solution is the concept of a {\it rigid quasilocal frame} (RQF). If we consider not the {\it three}-parameter family of volume-filling observers, but just the {\it two}-parameter family of observers on the boundary of the volume, who sweep out the three-dimensional Lorentzian spacetime manifold $\mathcal{B}$, the rigidity condition reduces to only {\it three} differential constraints on the same three functions, which always admit solutions, even in general relativity. Remarkably, the solutions are precisely analogous to the six-parameter family of rigid frames of Newtonian mechanics. They arise because the two-sphere boundary of any finite spatial volume (with the topology of a three-ball) always admits precisely six conformal Killing vector fields (three boosts and three rotations) that generate an action of the Lorentz group on the sphere~\cite{EMM2009,EMM2011}. Physically, an RQF is a congruence with zero expansion and shear, meaning the size and shape of the system boundary do not change, despite the fact that the boundary may be accelerating or rotating. This allows us to cleanly identify the most relevant energy and momentum fluxes crossing the system boundary (i.e., eliminate those fluxes due merely to changes in the size or shape of the boundary) and, moreover, to obtain simple, exact definitions for the elusive {\it gravitational} versions of those fluxes in terms of operationally-defined geometrical quantities on the boundary.

This paper is a fusion of these two solutions: Brown and York's quasilocal stress-energy-momentum tensor and our notion of an RQF. As we shall see, the real significance of the RQF approach is that it allows us to construct conservation laws for energy and momentum {\it without relying on any spacetime symmetries}. In reference~\cite{EMM2012} we constructed a completely general matter plus gravity RQF energy conservation law for spatially finite systems that does not rely on the existence of a timelike Killing vector field. In this paper we do the same, but for linear and angular momentum, without relying on the existence of a spacelike Killing vector field.

In {\S}\ref{Local} we begin with a local momentum conservation law based on the identity $\nabla_a ( T^{ab}\Psi_b ) = ( {\nabla_a T^{ab}} ) \Psi_b + T^{ab} \nabla_{(a} \Psi_{b)}$, where $\Psi^a$ is a spatial vector field determining the particular component of linear or angular momentum we are interested in. In general relativity, $\nabla_a T^{ab}=0$, and the local conservation law reduces to $\nabla_a ( T^{ab}\Psi_b ) = T^{ab} \nabla_{(a} \Psi_{b)}$. We argue that this differential conservation law integrated over a four-dimensional worldtube volume makes no physical sense unless two conditions are satisfied: (1) the frame (three-parameter bundle of worldlines) must be rigid in the sense discussed above, and (2) $\Psi^a$ must be a Killing vector field of the spatial three-metric on the quotient space of the geometrically rigid bundle of worldlines. Neither of these conditions is satisfied in general, and so such an integrated local momentum conservation law is not general. We argue that, ultimately, this failure results because the local approach, based on only the matter stress-energy-momentum tensor, $T^{ab}$, does not (and cannot) properly account for gravitational effects. As an example to close the section, we specialize this law to a stationary context and construct a general relativistic version of Archimedes' law. While Archimedes' law~\cite{Arch} forms the foundation of hydrostatics and has broad applications in a number of disciplines, its application in a general relativistic context has remained almost completely unexplored. We compare our general relativistic version of Archimedes' law to a similar law constructed by Eriksen and Gr{\o}n~\cite{Eriksen2006} in the context of accelerated observers in Rindler space (or equivalently, a uniform gravitational field).

In {\S}\ref{Quasilocal} we properly account for gravitational effects by replacing $T^{ab}$ with Brown and York's quasilocal matter plus gravity stress-energy-momentum tensor, and the identity $\nabla_a ( T^{ab}\Psi_b ) = ( {\nabla_a T^{ab}} ) \Psi_b + T^{ab} \nabla_{(a} \Psi_{b)}$ with an analogous identity defined in the boundary spacetime, $\mathcal{B}$ [see equation~(\ref{differential_quasilocal_conservation_law})]. The vector field $\Psi^a$ becomes a vector field $\psi^a$ tangent to $\mathcal{B}$. We argue that the integrated form of this differential conservation law always makes physical sense because: (1) the rigid quasilocal frames (RQFs) discussed above always exist, and (2) the six conformal Killing vector fields discussed above also always exist, and $\psi^a$ can always be taken to be one them (three boosts, corresponding to the three components of linear momentum, and three rotations, corresponding to the three components of angular momentum). Moreover, the resulting completely general matter plus gravity RQF momentum conservation law for spatially finite systems tells us something new about the physics of momentum conservation. Firstly, it reveals a simple, exact operational definition for gravitational momentum flux (mentioned above) that allows us to understand more deeply a wide variety of physical phenomena, including the simple example of a falling apple. Secondly, while both the local and quasilocal laws handle tangential (shear) stresses similarly, the quasilocal law treats stresses normal to the spatial boundary on a different footing, in a novel way that involves the quasilocal {\it pressure}, which can have both matter and gravitational (``geometrical") sources. We show that the quasilocal law reduces to the local law in the limit of a small-sphere RQF, but in general it involves completely new gravitational effects that are not accounted for in the local law. We close the section by deriving a quasilocal version of Archimedes' law which, unlike the local version we constructed at the end of {\S}\ref{Local}, is as general as such a law can be. We apply this law to an example of electrostatic weight and buoyant force in the context of the Reissner-Nordstr\"{o}m black hole. We present a brief summary and conclusions in {\S}\ref{Summary}.

\section{Local Momentum Conservation Law in General Relativity}\label{Local}

In this section we follow the standard {\it local} conservation law approach to construct an integrated momentum conservation law for matter fields in a finite volume of space in the context of general relativity. We argue that for this integrated law to make sense, physically, we must impose two conditions (a rigidity condition in time and a Killing vector condition in space) that {\it cannot} always be satisfied. When they {\it can} be satisfied, and when we can further specialize to a certain {\it stationary} context, we can construct a general relativistic version of Archimedes' law for matter fields (e.g., electromagnetism), which illustrates how Maxwell stress-like buoyant forces support the matter weight contained in a non-inertial reference frame. We compare this law with that constructed by Eriksen and Gr{\o}n~\cite{Eriksen2006} for electromagnetism in the context of uniformly accelerating (Rindler) observers in flat spacetime.

\subsection{General Analysis}\label{Local-1}

Given a smooth, four-dimensional Lorentzian manifold, $\cal M$, with metric $g_{ab}$ and associated derivative operator $\nabla_a$, the identity
\begin{equation}\label{differential_local_conservation_law}
\nabla_a (T^{ab}\Psi_b )=(\nabla_a T^{ab})\Psi_b + T^{ab}\nabla_{(a}\Psi_{b)}
\end{equation}
provides the differential form of a local conservation law for the current $T^{ab}\Psi_b$. Here $T^{ab}$ is the matter stress-energy-momentum tensor. In the context of general relativity, $\nabla_a T^{ab}=0$. To integrate this conservation law, we consider a spatially finite three-parameter family of observers with four-velocity vector field $u^a$ tangent to their congruence of worldlines. Let $\mathcal{B}$ denote the three-dimensional Lorentzian manifold boundary of this bundle of worldlines. Let $\Sigma_{i}$ and $\Sigma_{f}$ denote two finite spacelike three-surfaces slicing through the bundle (respectively the initial and final volumes of space), and $\Delta{\cal B}$ denote the section of $\mathcal{B}$ lying between $\Sigma_{i}$ and $\Sigma_{f}$. Finally, let $\Delta {\cal V}$ denote the finite spacetime four-volume contained within these boundaries. See Figure~1. Integrating equation~(\ref{differential_local_conservation_law}) over $\Delta {\cal V}$ yields the integrated form of this differential conservation law:
\begin{equation} \label{integrated_local_conservation_law}
\frac{1}{c} \int\limits_{\Sigma_f - \Sigma_i}  d{\Sigma}\,  T^{ab} u^\Sigma_a \Psi_b = \int\limits_{\Delta\mathcal{B}}  d \mathcal{B} \, T^{ab} n_a \Psi_b - \int\limits_{\Delta{\mathcal V}}  d{\mathcal V}\, T^{ab} \nabla_{(a} \Psi_{b)}.
\end{equation}
Here $\frac{1}{c}u^a_\Sigma$ denotes the timelike future-directed unit vector field orthogonal to $\Sigma_{i}$ and $\Sigma_{f}$, and $n^{a}$ denotes the spacelike outward-directed unit vector field orthogonal to $\mathcal{B}$.
\begin{figure}
\begin{center}
\includegraphics[scale=0.8]{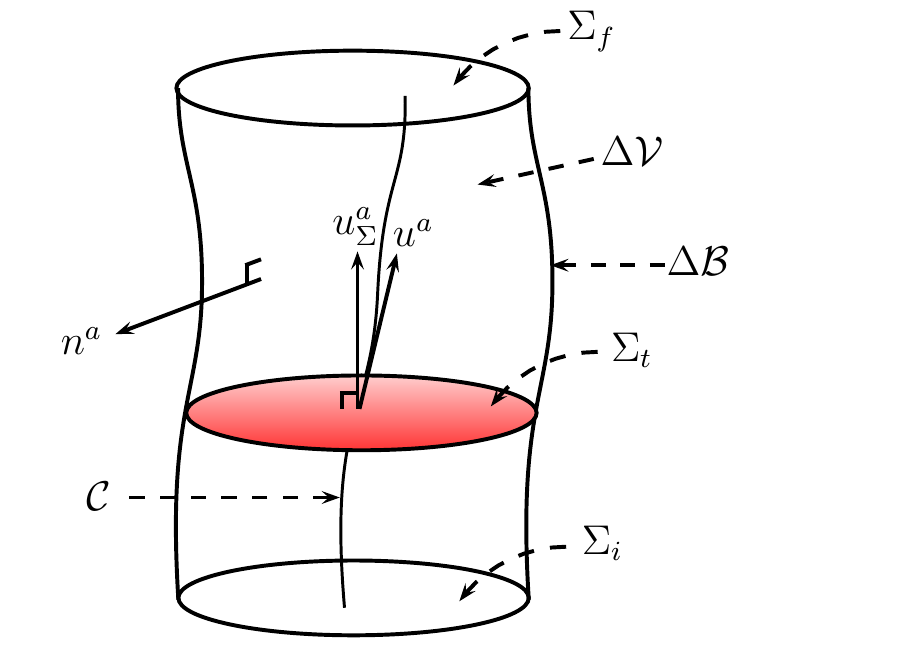}
\caption{An observer in the three-dimensional spatial volume $\Sigma_t$ follows a timelike worldline $\cal C$ with tangent four-velocity vector $u^a$, which is not necessarily parallel to the timelike vector field $u^a_{\Sigma}$ orthogonal to $\Sigma_t$.  The one-parameter family of spatial volumes, $\Sigma_t$, foliate the four-dimensional spacetime region, $\Delta \mathcal{V}$, whose timelike worldtube boundary, $\Delta\mathcal{B}$, has spacelike unit normal vector field $n^a$.}\label{LocalWorldtube}
\end{center}
\end{figure}
Roughly speaking, for a suitably-chosen spacelike vector field $\Psi^a$, this is a momentum conservation law that says that the difference in the matter three-momentum contained in the finite volumes $\Sigma_{i}$ and $\Sigma_{f}$ equals the flux of three-momentum that entered through the system boundary $\Delta\mathcal{B}$, plus a bulk $\Delta\mathcal{V}$ contribution arising when $\Psi^a$ is not a Killing vector field. We will be more precise later in this section.

To make the physical content of this law more transparent we decompose $T^{ab}$ as:
\begin{equation}\label{MatterSEM}
T^{ab} = \frac{1}{c^2}u^a u^b\mathbb{E}+2u^{(a}\mathbb{P}^{b)}-\mathbb{S}^{ab},
\end{equation}
where
\begin{align}
\mathbb{E} &= \frac{1}{c^2}u_a u_b T^{ab}=\frac{\rm Energy}{\rm Volume}=\frac{1}{8\pi}\left(E^2+B^2\right)\;\;{\rm (e.g.,\;electromagnetic\;energy\;density)}\nonumber\\
\mathbb{P}^a &= -\frac{1}{c^2}h^{a}_{\phantom{a}b}u_c T^{bc} = \frac{\rm Momentum}{\rm Volume}=\frac{1}{4\pi c}\epsilon^{a}_{\phantom{a}bc}E^b B^c\;\;{\rm (e.g.,\;Poynting\;vector\;over\;}c^2{\rm )}\label{Maxwell_stress}\\
\mathbb{S}^{ab} &= -h^{a}_{\phantom{a}c}h^{b}_{\phantom{b}d}T^{cd}=\frac{\rm Force}{\rm Area}=\frac{1}{4\pi}\left[E^a E^b+B^a B^b-\frac{1}{2}h^{ab}\left(E^2+B^2\right)\right]\;\;{\rm (e.g.,\; Maxwell\;stress)}.\nonumber
\end{align}
Here $h^a_{\phantom{a}b}=g^a_{\phantom{a}b}+\frac{1}{c^2}u^a u_b$ is the projection operator into the vector space orthogonal to the worldlines of the congruence, and $\epsilon_{abc}=\frac{1}{c}u^d\epsilon_{dabc}$ is the corresponding volume form in this space. The last equality corresponds to the example of electromagnetism, which will be used later when we compare our results with that in reference~\cite{Eriksen2006}. In this case, $E^a =  \frac{1}{c}F^{a b} u_b$ and $B^a = \frac{1}{2}\epsilon^{abc} F_{bc} $ are the proper electric and magnetic fields seen by the observers with four-velocity $u^a$~\cite{Wald1984}.

To get a momentum conservation law we set $\Psi^a = -\frac{1}{c}\Phi^a$, where $\Phi^a$ is orthogonal to (the worldlines of) the congruence. We then arbitrarily choose a time function on $\Delta{\mathcal V}$, i.e., a foliation of $\Delta{\mathcal V}$ by spacelike three-surfaces, $\Sigma_t$, of constant time parameter, $t$ (that coincide with $\Sigma_{i}$ and $\Sigma_{f}$ at times $t_i$ and $t_f$), and set $u^a=N^{-1}(\partial/\partial t)^a$, where $N$ is the lapse function. We naturally extend the definition of $u^a_\Sigma$ to all $\Sigma_t$ surfaces (as opposed to on just $\Sigma_{i}$ and $\Sigma_{f}$) as
\begin{equation}\label{definition_of_HSO_u}
u^a_\Sigma=\Gamma (u^a+V^a),
\end{equation}
where $V^a$ is orthogonal to the congruence, and $\Gamma=(1-V^2/c^2)^{-1/2}$ is a Lorentz factor. $V^a$ represents the three-velocity of fiducial observers who are `at rest' with respect to $\Sigma_t$ (whose hypersurface-orthogonal four-velocity is $u^a_\Sigma$) as measured by our congruence of observers (whose four-velocity is $u^a$). We will refer to these as the $u^a_\Sigma$- and $u^a$-observers, respectively. Note that while $u^a_\Sigma$ is hypersurface orthogonal, $u^a$ need not be, i.e., we are allowing for a twisting congruence. With these definitions, the conservation law in equation~(\ref{integrated_local_conservation_law}) reads:
\begin{align}\label{explicit_integrated_local_conservation_law}
\int\limits_{\Sigma_f - \Sigma_i}  d\hat{\Sigma} \, \left( \mathbb{P}^a+\frac{1}{c^2}\mathbb{S}^{ab}V_b\right)\Phi_a &=
\int\limits_{\Delta\mathcal{B}} N \, dt \, d\hat{\mathcal{S}}\; \mathbb{S}^{ab} n_a \Phi_b \nonumber\\
&- \int\limits_{\Delta{\mathcal V}}  N \, dt \, d\hat{\Sigma}\,  \left[\frac{1}{c^2}\mathbb{E}\,a^a\Phi_a +\mathbb{P}^a(\Theta_a^{\phantom{a}b}+\nu^c\epsilon_{ca}^{\phantom{ca}b} )\Phi_b-\mathbb{P}^a u^b \nabla_b \Phi_a +\mathbb{S}^{ab}\hat{\nabla}_{(a}\Phi_{b)} \right].
\end{align}
Here $d\hat{\Sigma}=\Gamma\,d\Sigma$ is the proper three-volume element seen by the $u^a$-observers. It is constructed from $h_{ab}$, the `radar ranging' metric that measures the orthogonal distance between neighboring worldlines of the congruence. Similarly, $d\hat{\mathcal{S}}$ is the proper two-surface element constructed from $\sigma_{ab}=h_{ab}-n_a n_b$, the `radar ranging' metric between neighboring worldlines of the congruence restricted to $\mathcal{B}$. In expanding the term $T^{ab} \nabla_{(a} \Psi_{b)}$ in equation~(\ref{integrated_local_conservation_law}) we made use of the following definitions associated with properties of the $u^a$-congruence: the observers' four-acceleration is defined as $a^a=u^b \nabla_b u^a$; the strain rate tensor (i.e., expansion and shear) of the congruence is defined as $\Theta_{ab}=h_{(a}^{\phantom{(a}c}h_{b)}^{\phantom{b)}d}\nabla_{c}u_{d}$, and the twist as $\nu_c=\frac{1}{2}\epsilon_{c}^{\phantom{c}ab}\nabla_{a}u_{b}$; and the derivative operator induced in the vector space orthogonal to the congruence is defined as $\hat{\nabla}_{a}\Phi_{b}=h_{a}^{\phantom{a}c}h_{b}^{\phantom{b}d}\nabla_{c}\Phi_{d}$.

The left-hand side of equation~(\ref{explicit_integrated_local_conservation_law}) is the change (between $\Sigma_i$ and $\Sigma_f$) in the $\Phi^a$-component of the matter (e.g., electromagnetic) momentum in the $\Sigma_t$ system, as measured by the $u^a$-observers. One might wonder why there is a Maxwell stress-like term in the integrand. Recall that the integrand started as $-\frac{1}{c^2}\,d{\Sigma}\,  T^{ab} u^\Sigma_a \Psi_b$, where $-\frac{1}{c^2} T^{ab} u^\Sigma_a$ is the matter four-momentum per unit volume as measured by the $u_\Sigma^a$-observers, who are co-moving with the $\Sigma_t$ system (i.e., their four-velocity is orthogonal to $\Sigma_t$). Multiplying by $d\Sigma$ gives the amount of matter four-momentum (again, as measured by the $u_\Sigma^a$-observers) contained in their proper volume element $d\Sigma$ of $\Sigma_t$. Contracting the resulting four-vector with $\Phi_b$ yields the $\Phi^a$-component of three-momentum as seen by the $u^a$-observers along {\it their} space axes. Finally, integrating over $\Sigma_t$ yields the total $\Phi^a$-momentum in the $\Sigma_t$ system, as measured by the $u^a$-observers at time $t$, who see the $\Sigma_t$ system as being in motion. So we are calculating the right thing. The $V^a$ in $u_\Sigma^a$ then results in the extra Maxwell stress-like term in the integrand.

Similarly, it is easy to see that instead of just $\mathbb{E}$ we will have $(\mathbb{E} - \mathbb{P}^a V_a)$ in the integrand when we choose $\Psi^a=\frac{1}{c}u^a$, and are dealing with an energy (instead of a three-momentum) conservation law~\cite{EMM2011}.\footnote{Note that we have changed the sign of the shift vector, $V^a$, from our previous papers to match the sign convention in references~\cite{Rohrlich1960,Jackson3rdEdition}.} This {\it covariant} definition of matter four-momentum is the same as Rohrlich's 1960 definition of electromagnetic four-momentum~\cite{Rohrlich1960}, which solved the infamous ``$4/3$ problem" in the classical theory of the electron (which was solved in essentially the same way by Fermi in 1922~\cite{Fermi1922}, but apparently forgotten). Compare Rohrlich's equation~(17) with our equation~(\ref{explicit_integrated_local_conservation_law}).\footnote{Alternatively, see section 16.5 in Jackson~\cite{Jackson3rdEdition}, in particular his equation~(16.44).} We are just seeing Rohrlich's electromagnetic four-momentum in special relativity generalized to arbitrary matter four-momentum in general relativity. To help clarify the parallel between our work here and Rohrlich's work on the electron, our $u^a$-observers (who will be moving {\it rigidly}---see next subsection) correspond to Rohrlich's {\it at rest} observers; they are observing a {\it moving} $\Sigma_t$ system that corresponds to Rohrlich's moving electron. Our $u_\Sigma^a$-observers correspond to the observers co-moving with Rohrlich's electron.

To understand the extra Maxwell stress-like term physically, consider for example a set of $u^a$-observers who see the $\Sigma_t$ system moving with velocity $V^a$ in the azimuthal direction (i.e., rotating relative to them), and choose $\Phi^a$ also in the azimuthal direction. Then the additional stress term is of the form $\frac{1}{c^2}\times {\rm Pressure}\times {\rm Velocity}$. In relativity theory, pressure makes a relativistic contribution to inertia; indeed, $\frac{1}{c^2}\times {\rm Pressure}$ has the dimensions of mass per unit volume. So matter (e.g., electromagnetic) pressure in a rotating system is equivalent to a rotating mass, which must contribute to the momentum ({\it angular} momentum in this example). Hence the $\frac{1}{c^2}\mathbb{S}^{ab}V_a\Phi_b$ term in equation~(\ref{explicit_integrated_local_conservation_law}).

So far, our analysis has been completely general. Before discussing the terms on the right-hand side of equation~(\ref{explicit_integrated_local_conservation_law}) it will be helpful to first simplify the equation by specializing it to the case of {\it rigid motion}, which, as we shall see, is ultimately necessary to achieve a physically sensible matter momentum conservation law for a finite-sized system.

\subsection{Specialization to Local Rigid Motion}\label{LocalRigid}

For reasons that will be made clear shortly, suppose that the $u^a$-observers are moving {\it rigidly}, i.e., that the orthogonal distance between the worldlines of all nearest neighboring pairs of $u^a$-observers is constant in time. This is equivalent to the condition $\Theta_{ab}=0$, i.e., we have a congruence with zero expansion and shear. Since this represents six differential constraints on three functions (the three independent components of $u^a$), this cannot always be realized. (However, as we will discuss in {\S}\ref{QuasilocalRigidMotion}, the quasilocal analogue of this condition {\it can} always be realized.) In the sequel we will simply assume we are in a context in which this rigidity condition {\it is} satisfied. As a first consequence, we obviously have that the spatial integration measures $d\hat{\Sigma}$ and $d\hat{\mathcal{S}}$ in equation~(\ref{explicit_integrated_local_conservation_law}) are time-independent.

Now let $\Upsilon^a$ denote any vector field orthogonal to the congruence, which the $u^a$-observers would consider to be purely spatial, i.e., to lie along their space axes. For this vector field to appear {\it stationary} to the rigidly-moving $u^a$-observers (i.e., not change with time), $\Upsilon^a$ must be Lie-dragged along the fibres, i.e., $\mathcal{L}_u \Upsilon^a \propto u^a$. Contracting both sides with $u_a$ reveals the proportionality factor, and we find that we require:
\begin{equation}\label{stationary_vectors}
\mathcal{L}_u \Upsilon^a = \frac{1}{c^2} (\Upsilon^b a_b) u^a.
\end{equation}
We will call spatial vector fields satisfying this condition {\it stationary}. In a context in which the rigidity condition holds, such stationary vector fields can be uniquely constructed throughout $\Delta\mathcal{V}$ given their specification on any one spatial three-surface, e.g., $\Sigma_i$.

Our conservation law is for the $\Phi^a$-component of the matter momentum. Obviously, we would certainly want $\Phi^a$ to be stationary, and will assume that such a choice has been made. Using equation~(\ref{stationary_vectors}) with  $\Upsilon^a =\Phi^a$ we find that $-\mathbb{P}^a u^b \nabla_b \Phi_a=\epsilon_{abc}\nu^a\mathbb{P}^b\Phi^c$, and so two of the terms in the $\Delta\mathcal{V}$ integrand in equation~(\ref{explicit_integrated_local_conservation_law}) can be combined into one:
\begin{equation}\label{Coriolis}
-\left[ \mathbb{P}^a\nu^c\epsilon_{ca}^{\phantom{ca}b}\Phi_b-\mathbb{P}^a u^b \nabla_b \Phi_a \right] = -2\epsilon_{abc}\nu^a\mathbb{P}^b\Phi^c.
\end{equation}
Recalling the usual vector formula for a Coriolis force, ${\bf F}_{\rm Cor}=-2\,m\,{\bf \Omega}\times {\bf v}$, equation~(\ref{Coriolis}) is clearly the $\Phi^a$-component of a matter Coriolis force density, which is associated with a change in the $\Phi^a$-component of the matter momentum as seen by our $u^a$-observers in the case that they are rotating (twisting congruence). Notice that it includes the correct factor of $-2$: half of the effect arises from the position-dependence of the relative velocity of points in the rotating frame (the first term on the left-hand side), and the other half arises from the rate at which the rotating frame coordinate axes change direction (the second term on the left-hand side). Relatedly, the $-\left[\frac{1}{c^2}\mathbb{E}\,a^a\Phi_a\right]$ term in equation~(\ref{explicit_integrated_local_conservation_law}) includes the $\Phi^a$-component of the sum of the matter Euler (``$-m\,{\bf \dot{\Omega}}\times {\bf r}$") and centrifugal (``$-m\,{\bf \Omega}\times {\bf \Omega}\times {\bf r}$") force densities, written in a covariant form that does not involve a radial vector (``${\bf r}")$.

Finally, we consider the term $\mathbb{S}^{ab}\hat{\nabla}_{(a}\Phi_{b)}$ in equation~(\ref{explicit_integrated_local_conservation_law}). If we introduce coordinates $x^I$, $I=1,\,2,\,3$, that label the worldlines of the congruence, then it is obvious that the rigidity condition $\Theta_{ab}=0$ is equivalent to $\dot{h}_{IJ}=0$, where $h_{IJ}$ are the spatial coordinate components of $h_{ab}$, and an over-dot denotes differentiation with respect to the parameter time, $t$. Assuming rigidity of the congruence, and stationarity of $\Phi^a$, a simple calculation shows that
\begin{equation}\label{KV_condition}
\hat{\nabla}_{(I}\Phi_{J)}=\frac{1}{2}\left(\Phi^K\partial_K h_{IJ}+2\,h_{K(I}\partial_{J)}\Phi^K\right),
\end{equation}
which are the only coordinate components of $\hat{\nabla}_{(a}\Phi_{b)}$ that do not identically vanish. Here $\partial_I$ denotes partial differentiation with respect to $x^I$. Recognizing the structure of a Lie derivative on the right-hand side of equation~(\ref{KV_condition}), requiring $\hat{\nabla}_{(a}\Phi_{b)}=0$ is thus equivalent to $\Phi^a$ being a Killing vector field with respect to $h_{ab}$. This is a natural condition to impose on $\Phi^a$ and $h_{ab}$; if it is not satisfied, it is not clear how meaningful it is to say we are dealing with the ``$\Phi^a$-component of the matter momentum" (more on this below). Of course this condition is not realizable in general. (However, we will see in {\S}\ref{QuasilocalRigidMotion} that the quasilocal analogue of this condition---a certain {\it conformal} Killing vector condition, {\it can} always be realized.) Here we will simply assume we are in a context in which this Killing vector condition {\it is} satisfied.

To summarize this subsection, assuming {\it rigid} motion, and a stationary {\it Killing} vector field $\Phi^a$ (Killing vector with respect to $h_{ab}$, not $g_{ab}$), equation~(\ref{explicit_integrated_local_conservation_law}) reduces to
\begin{equation}\label{rigid_KV_explicit_integrated_local_conservation_law}
\int\limits_{\Sigma_f - \Sigma_i}  d\hat{\Sigma} \, \left( \mathbb{P}^a+\frac{1}{c^2}\mathbb{S}^{ab}V_b\right)\Phi_a =
\int\limits_{\Delta\mathcal{B}} N \, dt \, d\hat{\mathcal{S}}\; \mathbb{S}^{ab} n_a \Phi_b
- \int\limits_{\Delta{\mathcal V}}  N \, dt \, d\hat{\Sigma}\,  \left[\frac{1}{c^2}\mathbb{E}\,a^a\Phi_a + 2\epsilon_{abc}\nu^a\mathbb{P}^b\Phi^c\right],
\end{equation}
where $d\hat{\Sigma}$ and $d\hat{\mathcal{S}}$ are time-independent. Within these assumptions, equation~(\ref{rigid_KV_explicit_integrated_local_conservation_law}) is a completely general matter momentum conservation law in the context of general relativity. It says that the change (between $\Sigma_i$ and $\Sigma_f$) in the $\Phi^a$-component of the matter momentum contained in the system, as measured by the $u^a$-observers, is equal to the $\Phi^a$-component of the impulse imparted to the system by matter stresses acting on the system boundary over that time interval (the $\Delta\mathcal{B}$ integral), plus a correction due to the non-inertial motion (acceleration and rotation) of the rigid $u^a$-frame (the $\Delta\mathcal{V}$ integral).

Before we move on, there is an important subtlety in equation~(\ref{rigid_KV_explicit_integrated_local_conservation_law}) worth pointing out. In the context of special relativity (and also Newtonian space-time), we sometimes come across the integral of a vector field---e.g., imagine equation~(\ref{rigid_KV_explicit_integrated_local_conservation_law}) without the contraction of the integrands with $\Phi_a$. This happens, for instance, in special relativity when we calculate the total electromagnetic force acting on the electromagnetic sources and fields in a given volume of space by integrating the Maxwell stress tensor (contracted with $n_a$) over the surface of the volume, i.e., an integral of the form $\int d\hat{\mathcal{S}}\; \mathbb{S}^{ab} n_a$. Of course such an integral makes no sense in the context of general relativity, where we cannot add vectors with different base points. But even when we contract the integrand with $\Phi_a$, so it makes mathematical sense, it won't make any physical sense unless $\Phi^a$ has certain special properties. In the context of equation~(\ref{rigid_KV_explicit_integrated_local_conservation_law}), $\Phi^a$ must somehow be uniquely determined throughout $\Delta\mathcal{V}$ by its value (and possibly a finite number of its derivatives) at a {\it single point} of $\Delta\mathcal{V}$, i.e., its degrees of freedom must be {\it discrete}, or `global', and in one-to-one correspondence with the degrees of freedom of the rigid frame itself. This, of course, is precisely the nature of a Killing vector field. In the present context, it is not difficult to show that the commutator of $h_{ab}$-compatible derivative operators acting on any {\it stationary} $\Phi^a$ depends on $\Phi^a$ at only a single point, i.e.,
\begin{equation}\label{Nature_of_Local_KV}
\hat{\nabla}_a\hat{\nabla}_b\Phi_c - \hat{\nabla}_b\hat{\nabla}_a\Phi_c = \hat{R}_{abc}^{\phantom{abc}d}\Phi_d,
\end{equation}
where $\hat{R}_{abc}^{\phantom{abc}d}$ is the Riemann tensor of $h_{ab}$, i.e., the curvature of the quotient space of our geometrically rigid bundle of worldlines. If we further demand that $\Phi^a$ be a Killing vector field with respect to $h_{ab}$ and define the antisymmetric object $A_{ab}=\hat{\nabla}_{[a}\Phi_{b]}$ it is not difficult to show that, for any vector field $X^a$ orthogonal to the congruence,
\begin{align}
X^a\hat{\nabla}_a\Phi_b &= X^a A_{ab} \label{IntegratingKV1}\\
X^a\hat{\nabla}_a A_{bc} &= -\hat{R}_{bca}^{\phantom{bca}d} X^a \Phi_d \label{IntegratingKV2}
\end{align}
in analogy with equations~(C.3.7-8) in reference~\cite{Wald1984}. Together with the stationarity condition on $\Phi^a$, this means $\Phi^a$ is uniquely determined throughout $\Delta\mathcal{V}$ by its value, and its antisymmetrized derivative, at a single point of $\Delta\mathcal{V}$.\footnote{In the case that $u^a$ is not hypersurface orthogonal one might worry about the integrability of equations~(\ref{IntegratingKV1}) and (\ref{IntegratingKV2}), i.e., their compatibility with the stationarity condition. However, a short calculation reveals that they are, in fact, compatible. If we start at a given point and integrate $\Phi^a$ to two different points on the same neighboring fibre (by following two different paths), the resulting $\Phi^a$ will satisfy equation~(\ref{stationary_vectors}).} In the case of our three-parameter family of worldlines, this integration data corresponds to six discrete degrees of freedom, resulting in six linearly independent Killing vector fields (three translational and three rotational) corresponding to six components of momentum (three linear and three angular) that can be analyzed in equation~(\ref{rigid_KV_explicit_integrated_local_conservation_law}). So requiring that $\Phi^a$ be a Killing vector field with respect to $h_{ab}$ is not only ``natural" (word used earlier), it is {\it crucial} for equation~(\ref{rigid_KV_explicit_integrated_local_conservation_law}) to make any physical sense (which, in turn, is predicated on the rigidity condition being satisfied). Indeed, $h_{ab}$ must admit the {\it maximal} number of Killing vector fields. The fact that this condition generically {\it cannot} be satisfied is a serious problem, not to mention the fact that the rigidity condition, too, generically {\it cannot} be satisfied. We emphasize this point because (as mentioned earlier), both the Killing (actually, {\it conformal} Killing) vector and rigidity conditions can always be realized in the quasilocal context, as we will discuss in {\S}\ref{QuasilocalRigidMotion}.

\subsection{Specialization to Local Archimedes' Law}\label{LocalArchimedes}

We now turn to formulating an Archimedes' law in general relativity. Archimedes' law is usually envisioned in a static, or at most a stationary context, which is certainly the simplest case, and the one we will explore here. But at this point, the rigid frame defined by the $u^a$-observers may still be undergoing time-dependent acceleration and/or rotation. So to formulate an Archimedes' law, it is natural to further assume that we are in a context in which $a^a$ and $\nu^a$ (which are both orthogonal to the congruence) are stationary. It is easy to show that the spatial coordinate components of $a_a$ and $\nu_a$ are:
\begin{align}
a_I &=\frac{1}{N}\dot{u}_I+c^2\partial_I\ln N\label{local_acceleration}\\
\nu_I &=\frac{1}{2}\epsilon_{I}^{\phantom{I}JK}\left(\partial_J u_K-\frac{1}{c^2}a_J u_K\right),\label{local_twist}
\end{align}
and the time coordinate components identically vanish. Assuming rigidity of the $u^a$-frame, stationarity of $a^a$ and $\nu^a$ is equivalent to  $\dot{a}_I=0$ and $\dot{\nu}_I=0$. Naturally, we assume that $V^a$---the three-velocity of the $u^a_\Sigma$-observers relative to the $u^a$-frame---is also stationary, i.e., $\dot{V}_I=0$. This is equivalent to $\dot{u}_I=0$ since equation~(\ref{definition_of_HSO_u}) implies $u_I=-V_I$. With $\dot{u}_I=0$, the acceleration $a_I =c^2\partial_I\ln N$ is a pure gradient. Demanding stationary acceleration then implies both a time-independent lapse function (so the {\it full} integration measures in equation~(\ref{rigid_KV_explicit_integrated_local_conservation_law}) are time-independent) and a stationary twist.

Having imposed conditions on the congruence of observers (rigidity, plus stationary acceleration and twist), and specified the $\Phi^a$ the observers will use (a stationary $h_{ab}$ Killing vector field), we have established a suitable, stationary framework in which the measurements of the matter (e.g., electromagnetic) field will be made. The final step is to assume that the matter field itself is stationary, which means $\mathbb{E}$, $\mathbb{P}^a$, and $\mathbb{S}^{ab}$ in equation~(\ref{Maxwell_stress}) are stationary. The left-hand side of equation~(\ref{rigid_KV_explicit_integrated_local_conservation_law}) then vanishes, and we are left with
\begin{equation}\label{local_Archimedes}
\int\limits_{\Delta\mathcal{B}} N \, dt \, d\hat{\mathcal{S}}\; \mathbb{S}^{ab} n_a \Phi_b=
\int\limits_{\Delta{\mathcal V}}  N \, dt \, d\hat{\Sigma}\,  \left[\frac{1}{c^2}\mathbb{E}\,a^a\Phi_a + 2\epsilon_{abc}\nu^a\mathbb{P}^b\Phi^c \right].
\end{equation}
This is an Archimedes' law for a general matter field in the context of general relativity, derived following the standard {\it local} conservation law approach. It basically says that the weight of the matter field in a non-inertial reference frame (right-hand side) is supported by a Maxwell stress-like buoyant force acting on the field inside through the boundary of the system (left-hand side).

In the case that the matter field is electromagnetism, this result is essentially identical to the main result, equation~(6.15), in Eriksen and Gr{\o}n's beautiful paper~\cite{Eriksen2006}, except for three key differences: (1) We have generalized the result from {\it static} to {\it stationary} contexts by allowing for stationary rotation (the addition of the Coriolis term). (2) We do not incorporate the lapse function, $N$, into the definition of the (electromagnetic) stress, energy, and momentum densities. For example, Eriksen and Gr{\o}n's definition of the Maxwell stress ($t_{ij}$ in their equation~(6.12)) in multiplied by $N$ ($g_0 x$ in their notation), compared to the standard definition in our equation~(\ref{Maxwell_stress}). We prefer to show the lapse function explicitly to highlight an important aspect of the actual {\it mechanism} of the buoyancy. In {\S}\ref{Example} we will exhibit a simple example in which $\int d\hat{\mathcal{S}}\; \mathbb{S}^{ab} n_a \Phi_b=0$, i.e., the Maxwell stress on one side of the surface balances that on other side, and so one might expect zero net buoyancy. That the net effect is {\it not} zero follows from the fact that the acceleration induces an inhomogeneous time dilation (encoded in $N$), so observers on one side of the surface experience the {\it same} (magnitude of) proper force, but for a {\it longer} proper time, than on the other side, resulting in a net buoyant impulse acting on the system. Similar consequences of an inhomogeneous time dilation were noticed in our earlier work~\cite{EMM2012}. (3) Eriksen and Gr{\o}n work in the context of Rindler observers in special relativity. Here we work in the context of accelerating and/or rotating rigid frames in general relativity. This is not to say, however, that equation~(\ref{local_Archimedes}) (or more generally, equation~(\ref{rigid_KV_explicit_integrated_local_conservation_law})) is describing what is really happening in the general relativistic context. We believe it is not, because it does not properly account for gravitational effects. To do so requires not a local, but a quasilocal approach, which we turn to presently.

\section{Quasilocal Momentum Conservation Law in General Relativity}\label{Quasilocal}

In this section we will use a quasilocal approach to construct an integrated momentum conservation law for both matter and gravitational fields contained in a finite volume of space. Such an approach was initiated by Brown and York~\cite{BY1993}, and we have further developed the approach in references~\cite{EMM2009,EMM2011,EMM2012}, in particular through the novel idea of a {\it rigid quasilocal frame} (RQF). Recall that the integrated local matter momentum conservation law in equation~(\ref{rigid_KV_explicit_integrated_local_conservation_law}) required two key conditions to make it physically sensible: a rigid motion condition and a Killing vector condition, neither of which can be satisfied in general. Here we show how these two serious obstacles, inherent in the local approach to constructing a momentum conservation law, are overcome in the quasilocal approach using RQFs, yielding a conservation law of general validity.  The essence of the solution is the proper inclusion of gravitational effects.

\subsection{General Analysis}\label{QuasilocalMomentumConservationLaw}

\begin{figure}
\begin{center}
\includegraphics[scale=0.8]{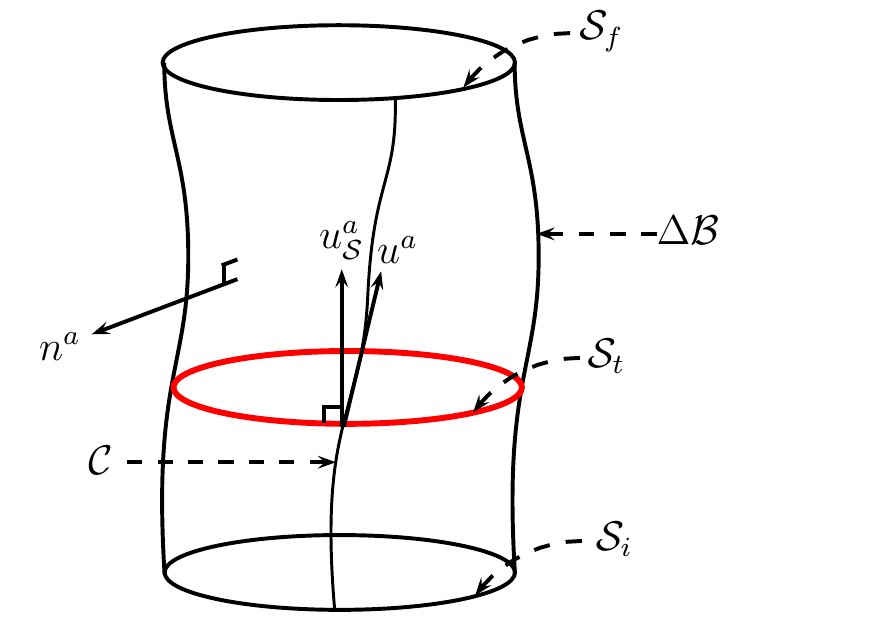}
\caption{An observer in the two-dimensional spatial surface $\mathcal{S}_t$ follows a timelike worldline $\cal C$ that lies in $\Delta\mathcal{B}$, and whose tangent four-velocity vector $u^a$ is not necessarily parallel to the timelike vector field $u^a_{\mathcal{S}}$ orthogonal to $\mathcal{S}_t$ and tangent to $\Delta\mathcal{B}$. The one-parameter family of spatial surfaces, $\mathcal{S}_t$, foliate the three-dimensional spacetime region, $\Delta \mathcal{B}$, whose boundaries are $\mathcal{S}_i$ and $\mathcal{S}_f$.}\label{QuasilocalWorldtube}
\end{center}
\end{figure}

We begin with an identity analogous to equation~(\ref{differential_local_conservation_law}), except constructed in the three-dimensional Lorentzian manifold, $\mathcal{B}$, the boundary of the bundle of worldlines defined in {\S}\ref{Local-1}, whose three-metric is $\gamma_{ab}=g_{ab}-n_a n_b$, with associated derivative operator $D_a$:
\begin{equation}\label{differential_quasilocal_conservation_law}
D_a (T_\mathcal{B}^{ab}\psi_b )=(D_a T_\mathcal{B}^{ab})\psi_b + T_\mathcal{B}^{ab}D_{(a}\psi_{b)}.
\end{equation}
Here $\psi^a$ is an arbitrary vector field tangent to $\mathcal{B}$, and the local matter stress-energy-momentum tensor, $T^{ab}$, in equation~(\ref{differential_local_conservation_law}) has been replaced with the quasilocal total (matter plus gravity) stress-energy-momentum tensor defined by Brown and York~\cite{BY1993} as $T_\mathcal{B}^{ab}=-\frac{1}{\kappa}\Pi^{ab}$. Here $\Pi_{ab}=K_{ab}-K\gamma_{ab}$ is the gravitational momentum canonically conjugate to $\gamma_{ab}$, $K_{ab}=\gamma_{a}^{\phantom{a}c}\nabla_c n_b$ is the extrinsic curvature of $\mathcal{B}$, and $\kappa=8\pi G/c^4$. The quasilocal analogue of integrating equation~(\ref{differential_local_conservation_law}) over $\Delta\mathcal{V}$ is integrating equation~(\ref{differential_quasilocal_conservation_law}) over $\Delta\mathcal{B}$. Denoting the boundaries of $\Delta\mathcal{B}$ as $\mathcal{S}_i$ and $\mathcal{S}_f$, the initial and final spacelike two-surfaces where $\Sigma_i$ and $\Sigma_f$ intersect $\mathcal{B}$, this integration yields:
\begin{equation} \label{integrated_quasilocal_conservation_law}
\frac{1}{c} \int\limits_{\mathcal{S}_f - \mathcal{S}_i}  d{\mathcal{S}}\,  T_{\mathcal B}^{ab} u^{\mathcal{S}}_a \psi_b = \int\limits_{\Delta\mathcal{B}}  d \mathcal{B} \, \left[ \frac{1}{\kappa} G^{ab} n_a \psi_b - T_{\mathcal B}^{ab} D_{(a} \psi_{b)} \right].
\end{equation}
See Figure~\ref{QuasilocalWorldtube}. Here $\frac{1}{c}u_{\mathcal{S}}^a$ denotes the timelike future-directed unit vector field tangent to $\mathcal{B}$ and orthogonal to $\mathcal{S}_i$ and $\mathcal{S}_f$, and we used the Gauss-Codazzi identity, $D_a \Pi^{ab}=n_a G^{ab}$, where $G^{ab}$ is the Einstein tensor associated with $g_{ab}$.

Note that this conservation law  is a purely geometrical identity relating the intrinsic and extrinsic geometry of $\mathcal{B}$ (through $D_a$ and $T_\mathcal{B}^{ab}$, respectively) to the geometry of the embedding space, $\mathcal{M}$ (through $G^{ab}$). The matter stress-energy-momentum tensor, initially absent in equation~(\ref{integrated_quasilocal_conservation_law}), will enter once  Einstein's equation is invoked, in which case the first term on the right-hand side becomes $\frac{1}{\kappa} G^{ab} n_a \psi_b=T^{ab} n_a \psi_b$. This term  looks very much like $T^{ab} n_a \Psi_b$, the matter stress term in the local conservation law in equation~(\ref{integrated_local_conservation_law}). However, there are two important differences. First, the origins of these two terms are completely different. Unlike $T^{ab} n_a \Psi_b$, $T^{ab} n_a \psi_b$ does {\it not} come from integrating a divergence. It comes from the fact that $D_a T_\mathcal{B}^{ab} \neq 0$ (in contrast to $\nabla_a T^{ab}=0$), and so is analogous to the Lorentz force density in electrodynamics (except that, unlike in electrodynamics, where the Lorentz force density acts only on the sources of the electromagnetic field, $D_a T_\mathcal{B}^{ab}$ acts on {\it all} of the fields in the system, matter and gravitational). The second important difference is that $\psi^a$ in $T^{ab} n_a \psi_b$ is {\it tangent} to $\mathcal{B}$, whereas $\Psi^a$ in $T^{ab} n_a \Psi_b$ need not be---it can (and often does) have a component in the normal direction, $n^a$. To see the significance of this, imagine a system with a two-sphere boundary, and we are interested in the `vertical', or `$Z^a$' component of the external matter force acting on the system (we will be more precise later). While there is no problem setting $\Psi^a=Z^a$ in the local conservation law, in the quasilocal law $\psi^a$ can accommodate only the {\it tangential} component of $Z^a$, not the normal component. So the quasilocal law seems to be missing the normal (pressure) contribution to the external matter force, $T^{ab} n_a n_b$. However, through an application of the `radial' Hamiltonian constraint of general relativity, we will show that this ``missing" normal matter force is found in the $T_{\mathcal B}^{ab} D_{(a} \psi_{b)}$ term in equation~(\ref{integrated_quasilocal_conservation_law}).

Before proceeding further with the analysis of equation~(\ref{integrated_quasilocal_conservation_law}), let us first introduce a more transparent notation for $T_{\mathcal B}^{ab}$. Following Brown and York~\cite{BY1993}, we resolve the quasilocal stress-energy-momentum tensor into components adapted to the $u^a$-observers:
\begin{equation}
T_\mathcal{B}^{ab} = \frac{1}{c^2}u^a u^b\mathcal{E}+2u^{(a}\mathcal{P}^{b)}-\mathcal{S}^{ab},
\end{equation}
where
\begin{equation}
\mathcal{E} = \frac{1}{c^2}u_a u_b T_\mathcal{B}^{ab}=\frac{\rm Energy}{\rm Area},\;\;
\mathcal{P}^a = -\frac{1}{c^2}\sigma^{a}_{\phantom{a}b}u_c T_\mathcal{B}^{bc}=\frac{\rm Momentum}{\rm Area},\;\;
\mathcal{S}^{ab} = -\sigma^{a}_{\phantom{a}c}\sigma^{b}_{\phantom{b}d}T_\mathcal{B}^{cd}=\frac{\rm Force}{\rm Length}.
\end{equation}
Here $\sigma^a_{\phantom{a}b}=g^a_{\phantom{a}b}+\frac{1}{c^2}u^a u_b-n^a n_b$ is the projection operator into the vector space tangent to $\mathcal{B}$ and orthogonal to the fibres of the congruence, and $\epsilon_{ab}=\frac{1}{c}u^c n^d\epsilon_{cdab}$ is the corresponding volume form in this space. These equations are exactly analogous to equations~(\ref{MatterSEM}) and (\ref{Maxwell_stress}), except
$\mathcal{E}$, $\mathcal{P}^a$, and $\mathcal{S}^{ab}$ refer to matter {\it and} gravity, whereas $\mathbb{E}$, $\mathbb{P}^a$, and $\mathbb{S}^{ab}$ refer to matter only (e.g., electromagnetism).\footnote{Our sign convention for the quasilocal stress is opposite to that of Brown and York~\cite{BY1993}, and our own previous work. Our new sign convention aligns with that of electromagnetism, and makes various arguments we will present in this section more physically sensible.}

Proceeding in analogy with the local case, to get a quasilocal momentum conservation law we set $\psi^a = -\frac{1}{c}\phi^a$, where $\phi^a$ is tangent to $\mathcal{B}$ and orthogonal to $u^a$. Letting $\mathcal{S}_t$ denote the intersection of $\Sigma_t$ with $\Delta{\mathcal B}$, we inherit from our local analysis an arbitrary time function on $\Delta{\mathcal B}$ (i.e., a foliation of $\Delta{\mathcal B}$ by spacelike two-surfaces, $\mathcal{S}_t$, which we assume are topologically two-spheres), and set $u^a=N^{-1}(\partial/\partial t)^a$ as before, where $N$ is the lapse function. In analogy to equation~(\ref{definition_of_HSO_u}), we extend the definition of $u^a_\mathcal{S}$ to all $\mathcal{S}_t$ surfaces as
\begin{equation}\label{definition_of_Boundary_HSO_u}
u^a_\mathcal{S}=\gamma (u^a+v^a),
\end{equation}
where $v^a$ is tangent to $\mathcal{B}$ and orthogonal to $u^a$, and $\gamma=(1-v^2/c^2)^{-1/2}$ is a Lorentz factor. Here $v^a$ represents the spatial two-velocity of fiducial observers who are `at rest' with respect to $\mathcal{S}_t$ (whose hypersurface-orthogonal four-velocity is $u^a_\mathcal{S}$) as measured by our congruence of observers (whose four-velocity is $u^a$). Note that, while $u^a_\mathcal{S}$ is hypersurface orthogonal, $u^a$ need not be. (Also note that, unlike $u_{\Sigma}^a$ appearing in equation~(\ref{definition_of_HSO_u}), $u_{\mathcal{S}}^a$ is independent of the choice of $\Sigma_t$ in the interior. In the quasilocal approach we are completely decoupled from the interior.) With these definitions, the conservation law in equation~(\ref{integrated_quasilocal_conservation_law}) reads:
\begin{align}\label{explicit_integrated_quasilocal_conservation_law}
\int\limits_{\mathcal{S}_f - \mathcal{S}_i}  d\hat{\mathcal{S}} \, \left( \mathcal{P}^a+\frac{1}{c^2}\mathcal{S}^{ab}v_b\right)\phi_a &=
-\int\limits_{\Delta\mathcal{B}} N \, dt \, d\hat{\mathcal{S}}\; T^{ab} n_a \phi_b \nonumber\\
&- \int\limits_{\Delta{\mathcal B}}  N \, dt \, d\hat{\mathcal{S}}\,  \left[\frac{1}{c^2}\mathcal{E}\,\alpha^a\phi_a +\mathcal{P}^a(\theta_a^{\phantom{a}b}+\nu\epsilon_{a}^{\phantom{a}b} )\phi_b-\mathcal{P}^a u^b D_b \phi_a +\mathcal{S}^{ab}\hat{D}_{(a}\phi_{b)} \right].
\end{align}
Analogous to $d\hat{\Sigma}=\Gamma\,d\Sigma$ in the local case, we have $d\hat{\mathcal{S}}=\gamma\,d\mathcal{S}$, the proper two-surface element seen by the $u^a$-observers on $\mathcal{B}$. In expanding the term $T_{\mathcal B}^{ab} D_{(a} \psi_{b)}$ in equation~(\ref{integrated_quasilocal_conservation_law}) we made use of the following definitions associated with properties of the $u^a$-congruence: the component of the observers' four-acceleration tangent to $\mathcal{B}$ is defined as $\alpha^a=\sigma^a_{\phantom{a}b} a^b$; the strain rate tensor (i.e., expansion and shear) of the congruence is defined as $\theta_{ab}=\sigma_{(a}^{\phantom{(a}c}\sigma_{b)}^{\phantom{b)}d} D_{c}u_{d}$, and the twist as $\nu=\frac{1}{2}\epsilon^{ab}D_{a}u_{b}$; and the derivative operator induced in the vector space tangent to $\mathcal{B}$ and orthogonal to the congruence is defined as $\hat{D}_{a}\phi_{b}=\sigma_{a}^{\phantom{a}c}\sigma_{b}^{\phantom{b}d}D_{c}\phi_{d}$.

Comparing this quasilocal momentum conservation law  with the local one in equation~(\ref{explicit_integrated_local_conservation_law}), we see that the left-hand side of both equations gives the change, between times $t_i$ and $t_f$, of the momentum contained in the system as measured by the $u^a$-observers, including the stress term required for relativistic covariance as discussed at the end of {\S}\ref{Local-1}. The two key differences are: (1) the momentum density in equation~(\ref{explicit_integrated_quasilocal_conservation_law}) is {\it quasilocal}---a momentum per unit area, versus per unit volume, and is meaningless unless it is integrated over the entire closed two-surface bounding a given volume; and (2) the quasilocal momentum density includes {\it all} contributions to the momentum in the system---matter plus gravity, versus matter only. Note that $\phi^a$ in $\mathcal{P}^a\phi_a$ is {\it tangent} to $\mathcal{B}$, whereas $\Phi^a$ in $\mathbb{P}^a\Phi_a$ need not be, so one might wonder if the quasilocal momentum density is missing a `normal' contribution to the momentum, similar to the missing normal component in $T^{ab}n_a\phi_b$ noted above. The answer is no. Unlike $T^{ab}$, $\mathcal{P}^a$ is {\it inherently quasilocal} in the sense that, to use the example introduced earlier, $\mathbb{P}^a Z_a$ corresponds to $\mathcal{P}^a\phi_a$ when $\phi^a$ is a conformal Killing vector on the two-sphere boundary representing a boost in the $Z^a$ direction. (We will be more precise below.)

As in the local approach, we will now simplify the right-hand side of equation~(\ref{explicit_integrated_quasilocal_conservation_law}) by specializing to the case of a reference frame in rigid motion, which admits a maximal set of spatial {\it conformal} Killing vector fields, $\phi^a$. Unlike in the local approach, this is {\it always} possible in the quasilocal approach.

\subsection{Specialization to Quasilocal Rigid Motion}\label{QuasilocalRigidMotion}

In a generic spacetime it is always possible to construct a {\it rigid quasilocal frame} (RQF), that is, a two-parameter family of $u^a$-observers comprising a Lorentzian manifold, $\mathcal{B}$, such that the orthogonal distance between the worldlines of all nearest neighboring pairs of $u^a$-observers is constant in time  ~\cite{EMM2009,EMM2011}. This is equivalent to the condition $\theta_{ab}=0$, i.e., a congruence with zero expansion and shear. Unlike the analogous condition in the local approach, this condition represents only {\it three} differential constraints on three functions (the three independent components of $u^a$), and can always be realized. In fact, the degrees of freedom remaining are precisely those of rigid frames in Newtonian space-time: three linear and three angular velocities, each with arbitrary time dependence~\cite{EMM2009,EMM2011}. Let us assume that our $u^a$-observers comprise such an RQF. As a first consequence, $d\hat{\mathcal{S}}$ in equation~(\ref{explicit_integrated_quasilocal_conservation_law}) is time-independent. Secondly, we are always able to construct spatial vector fields, $\upsilon^a$, in $\mathcal{B}$ (vector fields tangent to $\mathcal{B}$ and orthogonal to $u^a$) that are {\it stationary}: in analogy to equation~(\ref{stationary_vectors}), they satisfy $\mathcal{L}_u \upsilon^a = \frac{1}{c^2} (\upsilon^b \alpha_b) u^a$. Naturally, we will choose $\phi^a$ to be such a stationary spatial vector field, which is uniquely determined everywhere on $\mathcal{B}$ given its specification on any one two-surface, e.g., $\mathcal{S}_i$. As with equation~(\ref{Coriolis}), two of the terms in equation~(\ref{explicit_integrated_quasilocal_conservation_law}) then combine into one term:
\begin{equation}\label{Quasilocal_Coriolis}
-\left[ \mathcal{P}^a\nu\epsilon_{a}^{\phantom{a}b}\phi_b-\mathcal{P}^a u^b D_b \phi_a \right] = -2\nu \epsilon_{ab}\mathcal{P}^a\phi^b,
\end{equation}
which is the $\phi^a$-component of the {\it quasilocal} Coriolis force density of both the matter and gravitational fields in the system. Relatedly, the $-\left[\frac{1}{c^2}\mathcal{E}\alpha^a\phi_a\right]$ term in equation~(\ref{explicit_integrated_quasilocal_conservation_law}) includes the quasilocal versions of Euler and centrifugal force densities, again for both the matter and gravitational fields in the system.

Next, we consider the term $\mathcal{S}^{ab}\hat{D}_{(a}\phi_{b)}$ in equation~(\ref{explicit_integrated_quasilocal_conservation_law}).  Introducing coordinates $x^i$, $i=1,\,2$, that label the worldlines of the congruence, the rigidity condition $\theta_{ab}=0$ is equivalent to $\dot{\sigma}_{ij}=0$, where $\sigma_{ij}$ are the spatial coordinate components of $\sigma_{ab}$. Assuming rigidity of the congruence, and stationarity of $\phi^a$, we have, in analogy to equation~(\ref{KV_condition}),
\begin{equation}\label{quasilocal_KV_condition}
\hat{D}_{(i}\phi_{j)}=\frac{1}{2}\left(\phi^k\partial_k \sigma_{ij}+2\,\sigma_{k(i}\partial_{j)}\phi^k\right),
\end{equation}
which are the only coordinate components of $\hat{D}_{(a}\phi_{b)}$ that do not identically vanish. Here $\partial_i$ denotes partial differentiation with respect to $x^i$. Clearly, requiring $\hat{D}_{(a}\phi_{b)}=0$ is equivalent to $\phi^a$ being a Killing vector field of the `radar ranging' metric $\sigma_{ab}$. As discussed at the end of {\S}\ref{LocalRigid}, in the local approach we demanded that $\Phi^a$ be a Killing vector of the `radar ranging' metric $h_{ab}$ (i.e., $\hat{\nabla}_{(a}\Phi_{b)}=0$). There, this was the right thing to do, since a three-dimensional space admits at most six linearly independent Killing vectors, corresponding to three linear and three angular components of momentum. The problem was that such Killing vectors do not always exist. In the quasilocal approach, on the other hand, this would {\it not} be the right thing to do, since a two-dimensional space admits at most three Killing vectors, corresponding essentially to three components of angular momentum---we would be missing the the three components of linear momentum. We need a weaker condition.

The natural and appropriate condition turns out to be the {\it conformal} Killing vector (CKV) condition:
\begin{equation}\label{quasilocal_CKV_condition}
\hat{D}_{(a}\phi_{b)}=\frac{1}{2}\sigma_{ab}\hat{D}_c\phi^c.
\end{equation}
This condition represents two differential constraints on two functions of two variables, $\phi^i (x^j )$. Assuming the topology of $\mathcal{B}$ is $\mathbb{R}\times\mathbb{S}^2$ (as indicated earlier), then it is well-known that equation~(\ref{quasilocal_CKV_condition}) admits precisely six linearly independent CKV solutions, representing the action of the Lorentz group on the two-sphere (three boosts and three rotations). This is true regardless of $\sigma_{ab}$---the geometry of the quotient space of the rigid bundle, i.e., the size and shape of the topological two-sphere boundary of the system. It is this fact that gives RQFs the same six degrees of freedom as a rigid frame in Newtonian space-time, as mentioned above~\cite{EMM2009,EMM2011}. If we choose $\phi^a$ to be a boost (respectively, rotation) generator we are dealing with a linear (respectively, angular) momentum conservation equation.

Summarizing, in the quasilocal approach it is always possible to choose the $u^a$-observers to be in {\it rigid} motion (i.e., to comprise an RQF), and to choose a spatial vector field, $\phi^a$, that is a stationary conformal Killing vector field (CKV with respect to $\sigma_{ab}$, not $\gamma_{ab}$). Thus, equation~(\ref{explicit_integrated_quasilocal_conservation_law}) can always be reduced to the form:
\begin{equation}\label{reduced_integrated_quasilocal_conservation_law}
\int\limits_{\mathcal{S}_f - \mathcal{S}_i}  d\hat{\mathcal{S}} \, \left( \mathcal{P}^a+\frac{1}{c^2}\mathcal{S}^{ab}v_b\right)\phi_a =
\int\limits_{\Delta\mathcal{B}} N \, dt \, d\hat{\mathcal{S}} \left\{ \mathbb{S}^{ab} n_a \phi_b
- \left[\frac{1}{c^2}\mathcal{E}\,\alpha^a\phi_a +2\nu\epsilon_{ab}\mathcal{P}^a\phi^b + {\rm P}\,\hat{D}_a\phi^a \right]\right\},
\end{equation}
where ${\rm P}=\frac{1}{2}\sigma_{ab}\mathcal{S}^{ab}$ is the quasilocal {\it pressure} (force per unit {\it length}) between the worldlines of $\mathcal{B}$. Its physical interpretation will be discussed in the next subsection. Note that since $\phi^a$ is orthogonal to $u^a$ we may replace $-T^{ab} n_a \phi_b$ in equation~(\ref{explicit_integrated_quasilocal_conservation_law}) with $+\mathbb{S}^{ab} n_a \phi_b$, which emphasizes that this is a purely spatial stress term. This completely general RQF momentum conservation law for matter {\it and} gravitational fields is to be compared with the less general (valid only in special spacetimes that admit both a rigid three-parameter congruence and a spatial Killing vector field orthogonal to the congruence) and incomplete (includes matter fields only, not gravitational) local momentum conservation law in equation~(\ref{rigid_KV_explicit_integrated_local_conservation_law}).

There are two key differences between equations~(\ref{reduced_integrated_quasilocal_conservation_law}) and (\ref{rigid_KV_explicit_integrated_local_conservation_law}). The first is the obvious shift from integrations over volume densities to integrations over surface densities. To understand the physical significance of this shift, consider a simple problem in relativistic mechanics. Imagine being inside an accelerating box in flat spacetime that contains a freely-floating, massive object that appears to accelerate toward us; the object's momentum (relative to us) changes due to the acceleration of our frame. If the instantaneous proper acceleration of our frame is $a$, a simple calculation reveals that the change in momentum during an infinitesimal proper time interval $\Delta\tau$ is given by $\Delta p=-\frac{1}{c^2}E a \,\Delta\tau$, where $E$ is the instantaneous relativistic energy of the object (and the minus sign reflects the fact that $\Delta p$ and $a$ are in opposite directions). This explains the bulk term, $-\frac{1}{c^2}(\mathbb{E}\,d\hat{\Sigma})\,(a^a\Phi_a)\,(N\,dt)$, in the local law, equation~(\ref{rigid_KV_explicit_integrated_local_conservation_law}). But since momentum is a conserved quantity we now must ask: Where does this new momentum in the system come from? This might sound like a silly question---after all, momentum is frame-dependent, and we are just changing the frame! However, the question is perhaps not so silly when we ask it in the context of the equivalence principle. Instead of being inside an accelerating box, we could imagine that the box is at rest in a uniform gravitational field, and that the object is experiencing an acceleration toward us due to the ``force" of gravity. This ``force" acting over time imparts an impulse to the object, i.e., it represents a transfer of the momentum of {\it something} to the momentum of the object. If it was an electromagnetic force we would say that the ``something" is the electromagnetic field (and ultimately, the source of that field). Since the ``force" is gravitational, the ``something" must be the gravitational field (and ultimately, the source of that field). There must be some kind of surface flux representing {\it gravitational} momentum entering the box from the outside. According to the general relativistic equation~(\ref{reduced_integrated_quasilocal_conservation_law}), that gravitational momentum flux is $-\frac{1}{c^2}\mathcal{E}\,\alpha^a\phi_a$, which has dimensions of momentum per unit area per unit time. So the ``mass times acceleration" bulk term, $-\frac{1}{c^2}\mathbb{E}\,a^a\Phi_a$, in the local conservation law has become a bona fide gravitational momentum flux term, $-\frac{1}{c^2}\mathcal{E}\,\alpha^a\phi_a$, in the quasilocal conservation law, exactly as anticipated in our equivalence principle argument.

In the context of general relativity, the presence of mass-energy (matter or gravitational) inside a system causes a change in the spatial trace of the extrinsic curvature, $k=\sigma^{ab}K_{ab}$, of the two-sphere the RQF observers reside on. Since $\mathcal{E}=-k/\kappa$, measuring this change in extrinsic curvature (using a ruler) is operationally how the RQF observers measure the mass-energy in the system. For an everyday mass, $m$, and areal radius of the RQF, $r$, the magnitude of this change is exceedingly tiny, of order $Gm/c^2 r^2$ (as one can see by dimensional analysis). Nevertheless, multiplying this usually tiny general relativistic effect by the large number $c^2/G$ (and then integrating over the sphere) converts it into what we understand in classical mechanics as the mass of the object inside our frame of reference. This everyday mass, times an everyday acceleration, integrated over an everyday time interval, then gives an everyday change in momentum. But it is important to appreciate that the mass of an object times the acceleration of the frame in classical mechanics really represents a general relativistic transfer of gravitational momentum through the boundary of our reference frame, effected by a coupling between extrinsic curvature and acceleration. In other words, we claim that the simple expression, $-\frac{1}{c^2}\mathcal{E}\,\alpha^a\phi_a$, is actually the exact operational definition of gravitational momentum flux in general relativity (and similar comments apply to the $-2\nu\epsilon_{ab}\mathcal{P}^a\phi^b$ expression). So the RQF momentum conservation law in equation~(\ref{reduced_integrated_quasilocal_conservation_law}) leads us to a deeper understanding of physics: it explains in detail what is actually happening with regards to momentum transfer when, say, an apple falls. We discovered an exactly analogous result for energy transfer in reference~\cite{EMM2012}, except in place of a tiny change in extrinsic curvature we had a tiny frame dragging effect. But both are {\it general relativistic} effects.

The second key difference between the local and quasilocal momentum conservation laws is the extra ${\rm P}\,\mathcal{D}_a\phi^a$ term on the right-hand side of equation~(\ref{reduced_integrated_quasilocal_conservation_law}). In the next subsection we will show that it is this term that contains, in a subtle and interesting way, the missing normal ($-T^{ab} n_a n_b=+\mathbb{S}^{ab} n_a n_b$ pressure) contribution to the external matter force acting on the system, as alluded to earlier. (Remember that in equation~(\ref{reduced_integrated_quasilocal_conservation_law}), $-T^{ab}n_a\phi_b$ was replaced with $+\mathbb{S}^{ab}n_a\phi_b$, and so we are comparing $\mathbb{S}^{ab}n_a\Psi_b$ in the local law with $\mathbb{S}^{ab}n_a\psi_b$ in the quasilocal law. The {\it difference}, we claim, is contained in the extra ${\rm P}\,\mathcal{D}_a\phi^a$ term.)

\subsection{Role and Interpretation of the Quasilocal Pressure} \label{Interpretation}

General relativity imposes four constraint equations that intertwine the intrinsic and extrinsic geometry of $\mathcal{B}$ with matter. The RQF momentum conservation law in equation~(\ref{reduced_integrated_quasilocal_conservation_law}) represents two of these constraint equations. The RQF energy conservation law (see reference~\cite{EMM2012}) represents the third. The fourth constraint equation---the `radial' Hamiltonian constraint, originates in the geometrical identity:
\begin{equation}\label{RadialHamiltonianIdentity}
-2\,G_{ab}n^a n^b={}^\mathcal{B}\!R +\Pi_{ab}\Pi^{ab}-\frac{1}{2}\Pi^2,
\end{equation}
where ${}^\mathcal{B}\!R$ is the Ricci scalar of the intrinsic geometry of $\mathcal{B}$, $\Pi_{ab}=K_{ab}-K\gamma_{ab}$ (defined earlier) represents the extrinsic geometry of $\mathcal{B}$, and $\Pi=\gamma^{ab}\Pi_{ab}$. Substituting the Einstein equation ($G_{ab}=\kappa T_{ab}$) and the definition of the quasilocal stress-energy-momentum tensor ($\Pi_{ab}=-\kappa T^\mathcal{B}_{ab}$), we find that the radial Hamiltonian constraint is {\it linear} in the quasilocal pressure:
\begin{equation}\label{RadialHamiltonianPhysical}
2\kappa\, \mathbb{S}^{ab}n_a n_b={}^\mathcal{B}\!R +\kappa^2\left(\frac{1}{2}\mathcal{E}^2-2c^2\mathcal{P}^2+\tilde{\mathcal{S}}^2-2\mathcal{E}{\rm P}\right),
\end{equation}
where again we used the fact that $-T^{ab}n_a n_b=+\mathbb{S}^{ab}n_a n_b$. Here $\mathcal{P}^2=\mathcal{P}_a \mathcal{P}^a$ and $\tilde{\mathcal{S}}^2=\tilde{\mathcal{S}}_{ab} \tilde{\mathcal{S}}^{ab}$, where $\tilde{\mathcal{S}}_{ab}=\mathcal{S}_{ab}-{\rm P}\sigma_{ab}$ is the trace-free part of the quasilocal stress tensor. It is clear from this equation that there is a close relationship between the quasilocal pressure, $\rm P$, and the ``missing" external normal matter pressure, $\mathbb{S}^{ab}n_a n_b$, acting on the system. Solving for $\rm P$ we have:
\begin{equation}
{\rm P} = {\rm P}_{\rm mat} + {\rm P}_{\rm geom},\label{P_split}
\end{equation}
where we have split $\rm P$ into separate matter and geometry terms:
\begin{align}
{\rm P}_{\rm mat} &= -\frac{1}{\kappa\mathcal{E}}\, \mathbb{S}^{ab}n_a n_b,\label{MatterPressure}\\
{\rm P}_{\rm geom} &= \frac{1}{2\mathcal{E}}\left[ \frac{1}{\kappa^2} {}^\mathcal{B}\!R+\left( \frac{1}{2}\mathcal{E}^2 -2c^2\mathcal{P}^2+\tilde{\mathcal{S}}^2 \right) \right].\label{GeometricalPressure}
\end{align}

Before we can continue, it is important to appreciate that the quasilocal momentum conservation law in equation~(\ref{reduced_integrated_quasilocal_conservation_law}) includes gravitational effects, whereas the local momentum conservation law in equation~(\ref{rigid_KV_explicit_integrated_local_conservation_law}) does not. This means we can hope to meaningfully compare the two laws only in the limit where gravitational effects do not play a role, i.e., the limit of a small-sphere RQF, so that spacetime is nearly flat in the neighborhood of the RQF. For simplicity, we will take the RQF to be a round sphere of areal radius $r$, and construct series expansions of ${\rm P}_{\rm mat}$ and ${\rm P}_{\rm geom}$ in the first few leading powers of $r$. While this will get a bit messy, the messiness is only a result of unnaturally trying to cast a quasilocal law in the form of a local law; the quasilocal law itself is very simple and elegant.

For a small round-sphere RQF containing a smooth (non-singular) matter distribution, the quasilocal energy density has the general expansion:
\begin{equation}\label{QuasilocalEnergyExpansion}
\mathcal{E}=-\frac{2}{\kappa r} + \mathcal{E}_1\,r+\mathcal{O}(r^2).
\end{equation}
The dominant term for small $r$ comes from the fact, noted earlier, that $\mathcal{E}=-k/\kappa$, where $k=\sigma^{ab}K_{ab}$ is the spatial trace of the extrinsic curvature. For a round-sphere RQF of areal radius $r$ in flat spacetime, $k=2/r$. This leading term is often called the {\it vacuum energy density}; its role will be discussed in more detail at the end of this subsection. The next term in equation~(\ref{QuasilocalEnergyExpansion}), at order $r$, represents the lowest order at which matter can make a contribution to the energy of the system: when $\mathcal{E}$ is integrated over the two-sphere, this term becomes of order $r^3$, i.e., proportional to the spatial volume of the system. The lowest order at which gravity can make a contribution to the energy is order $r^3$ (i.e., a term of order $r^5$ when integrated over the two-sphere). We will verify both of these statements in the explicit example in {\S}\ref{Example}.

Substituting equation~(\ref{QuasilocalEnergyExpansion}) into equation~(\ref{MatterPressure}), and using the fact that for a non-singular matter distribution we must have $\mathbb{S}^{ab}n_a n_b =\mathcal{O}(1)$, we get
\begin{equation}\label{MatterPressureSeries}
{\rm P}_{\rm mat}=\frac{r}{2}\,\mathbb{S}^{ab}n_a n_b +\mathcal{O}(r^3 ).
\end{equation}
The factor of $r$ is obviously needed on dimensional grounds to convert a local pressure (force per unit area) into a quasilocal pressure (force per unit length). That the factor is precisely $r/2$ follows from a simple physical argument. Imagine that the round-sphere RQF is immersed in a matter field that is exerting a normal-normal stress $\mathbb{S}^{ab}n_a n_b$ that is {\it negative}, and for simplicity is uniform over the sphere. A negative $\mathbb{S}^{ab}n_a n_b$ corresponds to a local pressure pushing radially {\it inwards} on the surface of the sphere. The work done {\it by} the system (thought of in the local approach as the contents of the volume inside the sphere) in expanding the areal radius of the sphere from $r$ to $r+dr$ is then {\it positive}, and equal to $-4\pi r^2 \, \mathbb{S}^{ab}n_a n_b \, dr$. In the quasilocal approach, which doesn't ``know" anything about the contents of the volume of the sphere, the system is the surface of the sphere. For $\mathbb{S}^{ab}n_a n_b$ negative, we can imagine this surface to be like the elastic surface of a balloon, with an effective pressure, ${\rm P}_{\rm eff}$ (force per unit length), that is {\it negative}, i.e., the surface is under tension. Then the work done {\it by} the system against this tension in expanding the areal radius of the two-sphere from $r$ to $r+dr$ will be {\it positive}, and equal to $-{\rm P}_{\rm eff}\,d(4\pi r^2)=-8\pi r \, {\rm P}_{\rm eff}\,dr$. Equating the local and quasilocal expressions for the work done by the system, we have ${\rm P}_{\rm eff}=(r/2)\,\mathbb{S}^{ab}n_a n_b$, which explains the leading term in equation~(\ref{MatterPressureSeries}).

Returning to equation~(\ref{reduced_integrated_quasilocal_conservation_law}), we wish to show that the term $-{\rm P}_{\rm mat}\,\hat{D}_a\phi^a$ adds the correct normal matter pressure term to $\mathbb{S}^{ab} n_a \phi_b$. Using equation~(\ref{MatterPressureSeries}) we have
\begin{equation} \label{NormalMatterCorrection}
\mathbb{S}^{ab} n_a \phi_b -{\rm P}_{\rm mat}\,\hat{D}_a\phi^a = \mathbb{S}^{ab} n_a \Phi_b + \mathcal{O}(r^2),\;\;{\rm where}\;\; \Phi_b =\phi_b-\frac{r}{2}(\hat{D}_a\phi^a )n_b.
\end{equation}
The notation ``$\Phi_b$" is suggestive of the fact that $\Phi_b$  here can, indeed, be identified with the $\Phi_b$ appearing in the local momentum conservation law, equation~(\ref{rigid_KV_explicit_integrated_local_conservation_law}). For example, if we choose $\phi^a$ to be a CKV that generates a boost in the $Z$-direction, we can make $\Phi^a$ here equal the $Z^a$ spatial unit vector discussed in the second paragraph of {\S}\ref{QuasilocalMomentumConservationLaw}, with the usual components tangential and normal to the sphere. Introducing a spherical coordinate system adapted to the RQF, $x^a = (t,r,\theta,\phi)$, it is not difficult to show that we must have $\phi_a = (0,\mathcal{O}(r^2),-r\sin\theta,0)$, with $\hat{D}_a\phi^a =-(2/r)\cos\theta$---an exact result, and $n_a = (0,1+\mathcal{O}(r^2),0,0)$, which results in $Z_a = (0,\cos\theta+\mathcal{O}(r^2),-r\sin\theta,0)$, as required. Thus, we can write the right-hand side of our exact RQF momentum conservation law in equation~(\ref{reduced_integrated_quasilocal_conservation_law}) in the approximate form
\begin{equation}\label{reduced_integrated_quasilocal_conservation_law_series}
{\rm R.H.S.} =
\int\limits_{\Delta\mathcal{B}} N \, dt \, d\hat{\mathcal{S}} \left\{ \left[ \mathbb{S}^{ab} n_a \Phi_b +\mathcal{O}(r^2) \right]
- \left[\frac{1}{c^2}\mathcal{E}\,\alpha^a\phi_a +2\nu\epsilon_{ab}\mathcal{P}^a\phi^b +{\rm P}_{\rm geom}\,\hat{D}_a\phi^a \right]\right\}.
\end{equation}
Comparing with the right-hand side of the local momentum conservation law in equation~(\ref{rigid_KV_explicit_integrated_local_conservation_law}) we see that the matter stress terms are now {\it identical}, at least to the two leading orders in $r$.

But what of the ${\rm P}_{\rm geom}\,\hat{D}_a\phi^a$ term on the right-hand side? There is no such analogous term in the local momentum conservation law. What role does it play? In part, it serves to provide a `geometrical buoyant force' that supports (cancels) the dominant-in-$r$ vacuum weight coming from the vacuum energy density in the term $\frac{1}{c^2}\mathcal{E}\,\alpha^a\phi_a$. (Note that, since the vacuum energy density is {\it negative}, this weight is `up' and the buoyant force is `down'.) Of course such a cancelation {\it must} happen because there are no other terms in the conservation law at this order in $r$, and the conservation law is an identity. Nevertheless, it is instructive to see how this cancelation works in detail.

First, note that to lowest order in $r$, the lapse function must have the form
\begin{equation}\label{LapseSeries}
N=1+\frac{1}{c^2}\,{\bf A}\cdot {\bf r} +\mathcal{O}(r^2)=1-\frac{r^2}{2c^2}\hat{D}_a \alpha^a+\mathcal{O}(r^2),
\end{equation}
where ${\bf A}$ is the acceleration of a fiducial point at the center of the small round-sphere RQF in usual boldface vector notation. This will induce a tangential acceleration, $\alpha^a$, experienced by the RQF observers located on the surface of the sphere such that $\hat{D}_a \alpha^a=-(2/r^2)\,{\bf A}\cdot {\bf r}$ (as a simple calculation reveals); hence the alternative form of $N$, which will be more useful to us below. Thus we have, for the vacuum weight density:
\begin{equation}\label{VacuumWeight}
-N\,\frac{1}{c^2}\mathcal{E}\,\alpha^a\phi_a=+\frac{2}{\kappa c^2 r}\,\alpha^a\phi_a +\mathcal{O}(1),
\end{equation}
where $\alpha^a\phi_a=\mathcal{O}(1)$. What we need to show is that
\begin{equation}\label{VacuumBuoyancy}
-N\,{\rm P}_{\rm geom}\,\hat{D}_a\phi^a = -\frac{2}{\kappa c^2 r}\,\alpha^a\phi_a + \Omega +\mathcal{O}(1),
\end{equation}
where $\Omega$ represents a possible term of order $1/r$ or lower that integrates to zero over the sphere (i.e., is the divergence of a vector field). Note that to lowest order in $r$, the lapse function plays no role in equation~(\ref{VacuumWeight}), but it {\it will} play a critical role in equation~(\ref{VacuumBuoyancy}).

To prove equation~(\ref{VacuumBuoyancy}) we must examine the dominant terms in equation~(\ref{GeometricalPressure}). A bit of thought (and experience with RQFs) shows that the $\mathcal{P}^2$ and $\tilde{\mathcal{S}}^2$ terms will not play a dominant role, but the other two terms will. A short calculation reveals that, in the context of an RQF, the Ricci scalar associated with $\gamma_{ab}$ (the intrinsic geometry of $\mathcal{B}$) can be written as
\begin{equation}\label{RicciScalar}
{}^\mathcal{B}\!R=\hat{R}-\frac{2}{c^2}\,\left( \hat{D}_a\alpha^a + \frac{1}{c^2} \alpha_a\alpha^a -\nu^2 \right),
\end{equation}
where $\hat{R}$ is the Ricci scalar associated with $\sigma_{ab}$ (the geometry of the two-sphere quotient space), which, for the case of a round-sphere RQF of areal radius $r$, is $2/r^2$ (exact). The next-to-leading order term is $\hat{D}_a\alpha^a=\mathcal{O}(1/r)$, and the rest are higher order. Combining these results with the series expansion for $\mathcal{E}$ given in equation~(\ref{QuasilocalEnergyExpansion}) we find:
\begin{equation}\label{GeometricalPressureSeries}
{\rm P}_{\rm geom}=-\frac{1}{\kappa r}+\frac{r}{2\kappa c^2}\,\hat{D}_a\alpha^a +\mathcal{O}(r).
\end{equation}
The dominant term for small $r$ is a negative {\it vacuum pressure}, whose existence is intimately connected with the existence of the vacuum energy density through the geometrical identity $\mathcal{E}-2{\rm P}=(2/c^2\kappa)\,n_a a^a$ discussed more fully in reference~\cite{EMM2011}.\footnote{Recall that our sign convention for the quasilocal stress, and hence the quasilocal pressure, is opposite to that in our previous papers.} If we think of the negative vacuum pressure as a positive surface tension in the balloon analogy used earlier, then a similar calculation shows that the system must do an amount of positive work equal to $(8\pi/\kappa)\,dr$ to expand the areal radius by an amount $dr$, which is exactly the amount by which the (negative) vacuum energy of the system is reduced during this expansion. So the two vacuum entities are logically self-consistent.

Given that the vacuum pressure is {\it uniform} over the surface of the sphere (no gradient) one might assume that it does not create a buoyant force. However, when multiplied by the lapse function in equation~(\ref{LapseSeries}) it picks up a dipole cross term that accounts for precisely half of the proper time-integrated buoyant force (buoyant impulse) necessary to support the vacuum weight. (This is due to the acceleration-induced inhomogeneous time dilation mechanism mentioned in the last paragraph of {\S}\ref{LocalArchimedes}.) The next-to-leading order term in ${\rm P}_{\rm geom}$ is a dipole term that accounts for the other half of the buoyant force supporting the vacuum weight. Putting all of these results together we have:
\begin{equation}\label{VacuumBuoyancyActual}
-N\,{\rm P}_{\rm geom}\,\hat{D}_a\phi^a=\left[ \frac{1}{\kappa r}-\frac{r}{\kappa c^2} \, \hat{D}_a\alpha^a +\mathcal{O}(r) \right]\,\hat{D}_b\phi^b.
\end{equation}
Integrating $(\hat{D}_a\alpha^a)\,(\hat{D}_b\phi^b)$ by parts results in equation~(\ref{VacuumBuoyancy}) with $\Omega=\hat{D}_a\left[ (1/\kappa r)\,\phi^a-(r/\kappa c^2)\,\alpha^a\hat{D}_b\phi^b \right]$. Note that this part of the analysis has been purely geometrical. Matter contributions begin to appear in the quasilocal pressure only at order $r$---compare equations~(\ref{MatterPressureSeries}) and (\ref{GeometricalPressureSeries}).

Presumably we can continue the expansions started in equations~(\ref{VacuumWeight}) and (\ref{VacuumBuoyancy}) to the order in $r$ at which matter begins to contribute to the quasilocal weight density and check that, when the sum is integrated over the two-sphere, it gives the same result as the local matter weight density $-N\frac{1}{c^2}\mathbb{E}\,a^a\Phi_a$ in equation~(\ref{rigid_KV_explicit_integrated_local_conservation_law}) integrated over the three-volume inside the sphere. However, this would be prohibitively tedious (and besides, in the next subsection we will verify this explicitly in a nontrivial example). In any case, at this point it is clear that the quasilocal momentum conservation law in equation~(\ref{reduced_integrated_quasilocal_conservation_law}) is saying essentially the same thing as the local momentum conservation law in equation~(\ref{rigid_KV_explicit_integrated_local_conservation_law}), but in a novel way that has two key (and intimately related) advantages: (1) unlike in the local law, the rigidity and Killing vector-cum conformal Killing vector conditions can always be realized, and (2) unlike in the local law, the quasilocal law properly includes the effects of gravity. For a small-sphere RQF, the quasilocal law reduces to the local one; but as the system gets larger, the quasilocal law clearly departs from the local law in what it is saying.

Thus, the local law {\it does not explain what is really happening}. The local law treats the tangential (shear) components of force and weight on the same footing as the normal components. The quasilocal law does not. The normal components are accounted for in an entirely different way---through the quasilocal pressure term, $N\, {\rm P}\,\hat{D}_a\phi^a$, which has no analogue in the local law. When $\phi^a$ is a rotational CKV (so we are dealing with angular momentum), shear effects alone are sufficient, which is consistent with the fact that $\hat{D}_a\phi^a=0$ for a rotational CKV. But when $\phi^a$ is a boost CKV (so we are dealing with linear momentum), shear effects alone are {\it not} sufficient, which is consistent with the fact that $\hat{D}_a\phi^a$ is {\it not} zero in this case. The two-sphere integral of ${\rm P}\,\hat{D}_a\phi^a$ is then a measure of the corresponding $\ell=1$ spherical harmonic component of ${\rm P}$, i.e., the dipole component. A nontrivial dipole component represents a {\it gradient} in pressure, which, according to the quasilocal law, is how general relativity accounts for normal forces or weights. We examined only the dominant terms in equation~(\ref{GeometricalPressure}). There are clearly higher order nonlinear geometrical (gravitational) corrections that take us completely outside of the physics described by the local law.

As a final note, it is sometimes thought that the vacuum energy density in equation~(\ref{QuasilocalEnergyExpansion}) should be removed using the freedom in the definition of $T_\mathcal{B}^{ab}$ (interpreted as a freedom to choose the zero of energy for the system) to subtract a suitable reference stress-energy-momentum tensor (see, e.g., reference~\cite{BY1993}). However, we have seen here that the vacuum energy density and related vacuum pressure not only work together in a physically sensible way, but are actually {\it necessary} for the quasilocal momentum conservation law to have the correct small-sphere limit. The fact that $\mathcal{E}\rightarrow -2/\kappa r$ as $r\rightarrow 0$ is what allows the factor $-1/\kappa\mathcal{E}$ in equation~(\ref{MatterPressure}) to approach $r/2$ as $r\rightarrow 0$, which in turn leads to equation~(\ref{MatterPressureSeries}). This then allows $-{\rm P}_{\rm mat}\,\hat{D}_a\phi^a$ to add the correct normal matter pressure term to $\mathbb{S}^{ab} n_a \phi_b$ in equation~(\ref{NormalMatterCorrection}). Without this mechanism, we would have only a shear matter stress, and the quasilocal law would not reduce to the local law in the small-sphere limit. But this leads to a potential problem: a seemingly nonphysical dominant-in-$r$ vacuum weight coming from the term $\frac{1}{c^2}\mathcal{E}\,\alpha^a\phi_a$ in equation~(\ref{reduced_integrated_quasilocal_conservation_law}). As we have just seen, what comes to the rescue is the geometrical pressure in equation~(\ref{GeometricalPressure}), in particular the vacuum pressure plus the first subleading term, i.e., the two leading terms in $\rm P$ before matter begins to contribute. This suggests that perhaps the vacuum energy (and attendant vacuum pressure) cannot be freely removed, and might actually be physically real in some sense.

\subsection{Specialization to Quasilocal Archimedes' Law with Example}\label{Example}

We begin with the completely general RQF momentum conservation law for matter and gravitational fields in a finite volume given in equation~(\ref{reduced_integrated_quasilocal_conservation_law}), and specialize to a stationary context suitable for Archimedes' law. As in {\S}\ref{LocalArchimedes}, the first step is to assume certain properties of the frame of reference, namely, that the RQF observers are experiencing at most {\it stationary} acceleration and twist. In the RQF-adapted coordinates $x^a=(t,r,x^i )$ introduced earlier (where $x^i$ label the two-parameter family of worldlines), the spatial coordinate components of acceleration and twist are
\begin{align}
\alpha_i &=\frac{1}{N}\dot{u}_i+c^2\partial_i\ln N\label{quasilocal_acceleration}\\
\nu &=\frac{1}{2}\epsilon^{ij}\left(\partial_i u_j-\frac{1}{c^2}\alpha_i u_j\right);\label{quasilocal_twist}
\end{align}
compare with equations~(\ref{local_acceleration}) and (\ref{local_twist}). In further analogy to {\S}\ref{LocalArchimedes}, assuming also that $v^a$ (the two-velocity of the fiducial $u^a_\mathcal{S}$-observers relative to the $u^a$-frame) is also stationary implies that $\dot{u}_i=0$, in which case demanding stationary acceleration ($\dot{\alpha}_i=0$) implies that the lapse function (and thus the full integration measure) is time-independent and, in turn, the twist is stationary ($\dot{\nu}=0$). Recall also that the component of momentum the observers are measuring is specified by a {\it stationary} CKV, $\phi^a$. The second step is to assume properties of the matter and gravitational fields themselves, namely, that $\mathcal{E}$, $\mathcal{P}^a$ and $\mathcal{S}^{ab}$ are all stationary. The left-hand side of equation~(\ref{reduced_integrated_quasilocal_conservation_law}) then vanishes (the momentum in the system, if any, does not change with time) and we are left with
\begin{equation}
\int\limits_{\Delta\mathcal{B}} N \, dt \, d\hat{\mathcal{S}} \left[ \mathbb{S}^{ab} n_a \phi_b - {\rm P}_{\rm mat}\,\hat{D}_a\phi^a\right] =
\int\limits_{\Delta\mathcal{B}} N \, dt \, d\hat{\mathcal{S}} \left[\frac{1}{c^2}\mathcal{E}\,\alpha^a\phi_a +2\nu\epsilon_{ab}\mathcal{P}^a\phi^b +{\rm P}_{\rm geom}\,\hat{D}_a\phi^a \right],\label{quasilocal_Archimedes}
\end{equation}
where we used equation~(\ref{P_split}) and rearranged some terms.

We contend that equation (\ref{quasilocal_Archimedes}) is the fully general relativistic analogue of Archimedes' law for matter and gravitational fields contained in a finite volume.  It says essentially the same thing as the local Archimedes' law in equation~(\ref{local_Archimedes}), but does not rely on any spacetime symmetries.  To achieve this generality, the bulk integral on the right-hand side is replaced with a surface integral, and there are additional quasilocal pressure terms that account for normal forces and weights in a general relativistically correct way. Roughly speaking, it says that the weight of the matter and gravitational mass-energy in a non-inertial reference frame (right-hand side) is supported by a Maxwell stress-like buoyant force acting on the fields inside the system through the boundary of the system (left-hand side).

\begin{figure}
\begin{center}
\includegraphics[scale=0.6]{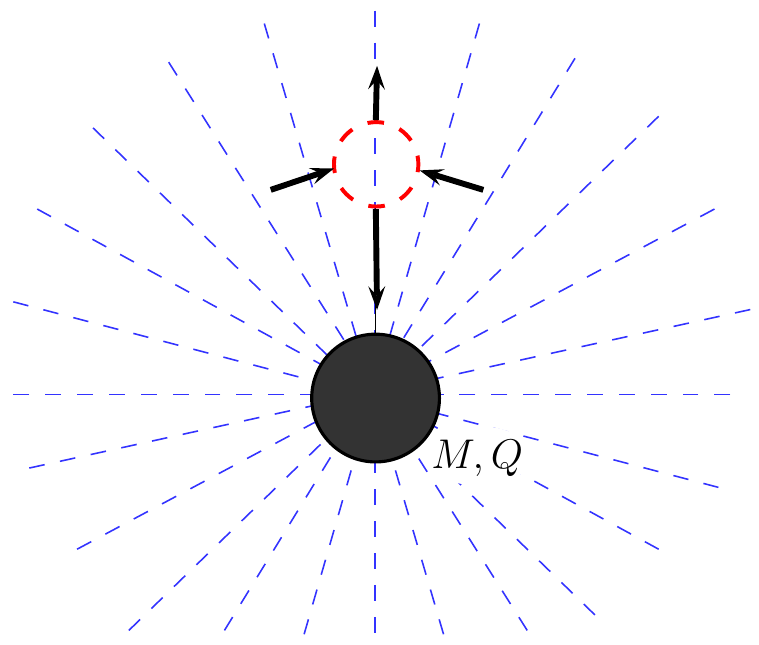}
\caption{The weight of the electrostatic field in a volume of space a fixed distance from a Reissner-Nordstr\"{o}m black hole is supported by the tensions along the electric field lines and pressures orthogonal to them at the boundary of the volume. For this to be true one might expect that the electric field lines exert a net upward force. Actually, they do not. Instead, there is a net buoyant {\it impulse} arising from an acceleration-induced inhomogeneous time dilation that enhances the upward impulses relative to the downward ones.}\label{RQFinRN}
\end{center}
\end{figure}

As a concrete example of this Archimedes' law, and one that is a close general relativistic analogue of the special relativistic work done in reference~\cite{Eriksen2006}, we consider a small round-sphere RQF hovering a fixed distance above a Reissner-Nordstr\"{o}m black hole. See Figure~\ref{RQFinRN}. In this paragraph we give a brief description of the physical mechanism behind Archimedes' law in this example; below we give the mathematical details of the analysis. We assume that the round-sphere RQF has an areal radius $r$, and conduct the analysis in powers of $r$ far enough to include the first two terms in the energy density expansion given in equation~(\ref{QuasilocalEnergyExpansion}), i.e., enough to be able to calculate the gravitational vacuum energy and lowest order electrostatic energy contained inside the sphere. Operationally, the RQF observers measure this energy by measuring (with a ruler) its local effect on the extrinsic geometry of the sphere, and then integrate this local effect over the entire sphere. Moreover, by the fact that their local gyroscopes do not precess, they conclude that there is zero momentum in the system.\footnote{In reference~\cite{EMM2012} we discuss how mass-energy in motion produces a relatively tiny frame dragging effect that causes the RQF observers' local gyroscopes to precess. Roughly speaking, rotating this relatively tiny precession rate vector through 90 degrees and multiplying by the large number $c^2/8\pi G$ yields the quasilocal momentum density, $\mathcal{P}^a$, which when integrated over the sphere gives what we understand in classical mechanics as the momentum of the mass-energy in motion.} On the other hand, using local accelerometers, the RQF observers measure the proper acceleration required for them to hover a fixed distance above the black hole. Thus seeing that their (rigid quasilocal) frame of reference is accelerating, but that the momentum of the mass-energy inside is not changing (it is always zero), the RQF observers conclude that there must be a Maxwell stress at the surface of the sphere acting on the electrostatic field inside, causing it to accelerate along with their frame. By measuring the local electrostatic field they are at first surprised to see that this Maxwell stress actually averages to {\it zero} over the surface of the sphere. It produces no net buoyant force. But then they realize that, because of their acceleration (or equivalently, the spacetime curvature), proper time elapses at a greater rate at the top of the sphere (the point furthest from the black hole) than at the bottom, and this results in a nonzero net buoyant {\it impulse}. This important effect of an acceleration-induced inhomogeneous proper time was mentioned earlier, and will be discussed in more detail below.\footnote{Incidentally, the quasilocal pressure in equation~(\ref{reduced_integrated_quasilocal_conservation_law}) can be measured indirectly using the identity $2{\rm P}=\mathcal{E} - (2/c^2\kappa)n_a a^a$ (discussed in reference~\cite{EMM2011}), i.e., operationally, {\rm P} is essentially the difference between the quasilocal energy density (measured with a ruler) and the normal component of proper acceleration (measured with an accelerometer). The split between ${\rm P}_{\rm mat}$ and ${\rm P}_{\rm geom}$ in equation~(\ref{quasilocal_Archimedes}) is then determined by measuring ${\rm P}_{\rm mat}$ through its relation to the normal matter pressure---see equation~(\ref{MatterPressure}). Alternatively, and probably closer to the spirit of the quasilocal approach, we would move the ${\rm P}_{\rm mat}$ term in equation~(\ref{quasilocal_Archimedes}) to the right-hand side and deal with the full quasilocal pressure, $\rm P$.}

We begin with the line element of the Reissner-Nordstr\"{o}m (RN) black hole in quasi-Cartesian coordinates $X^a = (X^0, X^I) = (cT, X,Y,Z),\, I=1,2,3$:
\begin{align}\label{RNCartesian}
ds^2 = - F^2 (R) c^2 dT^2 + \delta_{IJ} dX^I dX^J + \frac{1}{R^2} \left( \frac{1}{F^2 (R)} - 1 \right) \delta_{IK} \delta_{JL} X^I X^J dX^K dX^L ,
\end{align}
where $\delta_{IJ}$ is the Kronecker delta function, $R^2 = X^2 + Y^2 + Z^2$, $F(R) = \sqrt{1 - \frac{2GM}{c^2R} + \frac{G Q^2}{c^4 R^2}}$, and $M$ and $Q$ are the mass and charge of the black hole, respectively. We then construct a small spherical RQF about the point $Z=R_0$ on the $Z$-axis via the coordinate transformation
\begin{align}\label{RQF_coordinate_embedding}
T = \frac{1}{F(R_0)}\, t, \qquad X = r \sin \theta \cos \phi, \qquad Y = r \sin \theta \sin \phi, \qquad Z = R_0 + \sum^{\infty}_{n=1} f_{n} (\theta) \, r^n,
\end{align}
where $(t,r,x^i )$ are the RQF-adapted coordinates. The angular coordinates $x^i=(\theta,\phi)$ label the two-parameter family of RQF observers on the sphere (and their worldlines in the congruence, $\mathcal{B}$). The radial coordinate, $r\ll R_0$, is a small parameter related to the size of the sphere. The time coordinate, $t$, labels a natural choice of two-spheres, $\mathcal{S}_t$, foliating $\mathcal{B}$. In the case of flat spacetime, i.e., $F(R)=1$, the choice of functions $f_1(\theta)=\cos\theta$ and $f_{n>1}(\theta)=0$ represents a set of RQF observers sitting on a round sphere of areal radius $r$, with standard spherical coordinates $(\theta,\phi)$, centered on a point along the $Z$-axis a distance $R_0$ from the origin. This constitutes an RQF with two-surface metric $\sigma_{ij} = r^2 \mathbb{S}_{ij}$, where $\mathbb{S}_{ij}$ is the standard spherical coordinate metric on a unit two-sphere. As we turn on the mass and charge of the RN black hole, i.e., $F(R)\neq 1$, we would like the RQF observers to maintain the {\it same} round sphere two-surface metric ($\sigma_{ij} = r^2 \mathbb{S}_{ij}$, for which $d\hat{\mathcal{S}}=r^2\sin\theta\,d\theta\,d\phi$), but the presence of spacetime curvature means that in the RQF coordinate embedding in equation~(\ref{RQF_coordinate_embedding}) we will have to rescale $f_1(\theta)$, and will need an infinite number of higher order corrections, $f_{n>1}(\theta)\neq 0$.\footnote{Note that the functions $f_n(\theta)$ do not depend on $\phi$ owing to the symmetry of the RN spacetime under rotations about the $Z$-axis.}
Using GRTensorII running under Maple, we have solved for these corrections up to order $r^4$. At the two lowest orders we find:\footnote{Note that at each order the functions $f_n(\theta)$ are determined only up to a constant of integration by the round-sphere RQF conditions (being differential in nature). We fix this residual gauge freedom so as to recover the flat spacetime result as $r\rightarrow 0$. For example, one actually finds $f_1 (\theta) = F(R_0) \cos \theta + C_1$, but we take $C_1 = 0$ to make $g_{rr} \rightarrow 1$ as $r\rightarrow 0$. Also note that we rescale $t$ by $1/F(R_0)$ in equation~(\ref{RQF_coordinate_embedding}) so that $g_{tt}\rightarrow -c^2$ as $r\rightarrow 0$.}
\begin{align}
f_1 (\theta) &= F(R_0) \cos \theta, \label{f1} \\
f_2(\theta) &= \frac{1}{4}\left[ F(R_0)F^\prime (R_0) + \frac{1-F^2(R_0)}{R_0} \right](2\cos^2\theta-1)+\frac{1}{4}\left[ F(R_0)F^\prime (R_0) - \frac{1-F^2(R_0)}{R_0} \right],\label{f2}
\end{align}
where $F^\prime(R_0)=dF(R_0)/dR_0$. The rescaling of $f_1 (\theta)$ by $F(R_0)$ represents a uniform contraction of the embedded $r={\rm constant}$ coordinate sphere along the $Z$-axis (but not the $X$- or $Y$-axes) to account for the fact that the spatial line element at the center of the sphere is $ds^2=dX^2+dY^2+dZ^2 / F^2 (R_0)$, i.e., proper distance along the $Z$-axis at $R=R_0$ is given by $ds = dZ / F(R_0)$; the coordinate contraction thus maintains a geometrically round sphere of areal radius $r$. The correction at next order in $r$, $f_2(\theta)$, which goes to zero in the flat spactime limit $F(R) = 1$, includes monopole and quadrupole terms, and in general there are an infinite number of higher order multipole corrections at higher order in $r$.

Our Maple calculation reveals that the quasilocal energy density is given by
\begin{equation}\label{script_E_RN}
\mathcal{E} = - \frac{2}{\kappa r} + \left[ \frac{1}{3}\mathbb{E}_\mathtt{EM} + \left( \frac{M c^2}{8\pi R_0^3} - \frac{2}{3}\mathbb{E}_\mathtt{EM} \right)\left( 3 \cos^2 \theta - 1 \right) \right] r
 + \mathcal{O}(r^2)
\end{equation}
where $\mathbb{E}_\mathtt{EM}=|\vec{E}_0|^2/8\pi$ is the energy density (energy per unit {\it volume}) of the electrostatic field, evaluated at the center of the sphere, and $|\vec{E}_0|=Q/R_0^2$ is the magnitude of the electrostatic field, also evaluated at the center. Observe that the quasilocal energy density has the general form proposed in equation (\ref{QuasilocalEnergyExpansion}). Multiplying $\mathbb{E}_\mathtt{EM}$ by $r/3$ converts the electrostatic volume energy density to an effective surface energy density (energy per unit area). Upon integrating $\mathcal{E}$ over the surface of the sphere, the $\ell=2$ spherical harmonic term makes no net contribution and we find that the total energy inside the RQF is
\begin{equation}\label{TotalEnergy}
E = - \frac{8 \pi}{\kappa} r + \mathbb{E}_\mathtt{EM}\left(\frac{4\pi}{3} r^3\right)+ \mathcal{O}(r^5).
\end{equation}
The first term on the right-hand side is the (negative) vacuum energy, and the second is the electrostatic field energy (at lowest order in $r$). Note: the Maple calculation shows that the order $r^2$ term in equation~(\ref{script_E_RN}) is actually a sum of $\ell=1$ and $\ell=3$ spherical harmonics that integrate to zero, so there is no order $r^4$ contribution to equation~(\ref{TotalEnergy}). General relativistic curvature effects begin to appear only at order $r^5$, as mentioned earlier.

Next, we compute the radial electrostatic pressure exerted on the field inside the RQF (normal-normal component of the Maxwell stress tensor) and find
\begin{equation} \label{TnnRN}
\mathbb{S}^{ab} n_{a} n_{b} = \mathbb{E}_\mathtt{EM}\, (2\cos^2\theta-1)-4\,\mathbb{E}_\mathtt{EM}\,\frac{F(R_0)}{ R_0}\, (3\cos^3\theta-2\cos\theta) \,r+ \mathcal{O}(r^2)
\end{equation}
Observe that at zeroth order in $r$ this pressure is directed radially outwards near the poles of the sphere (where the electric field lines are mainly orthogonal to the surface, and under tension) and inwards near the equator (where the electric field lines are mainly parallel to the surface, and repel each other)---see Figure~\ref{RQFinRN}. The correction at order $r$ accounts for the non-uniform (radial) nature of the electrostatic field around the black hole, and reduces to the special relativistic result when $F(R)=1$. Observe also that $\mathbb{S}^{ab}n_a n_b =\mathcal{O}(1)$, which is required for equations~(\ref{MatterPressureSeries}) and (\ref{NormalMatterCorrection}) to be valid. We have similarly calculated $\mathbb{S}^{ab} n_{a} \phi_{b}$, the electrostatic shear stress exerted on the field inside the RQF. Substituting equation~(\ref{TnnRN}) into equation~(\ref{MatterPressureSeries}), and recalling that $\hat{D}_a\phi^a =-(2/r)\cos\theta$ (an exact result), we find that the integrand on the left-hand side of equation~(\ref{quasilocal_Archimedes}) is given by
\begin{equation}\label{TnPhiRN}
\mathbb{S}^{ab} n_a \phi_b - {\rm P}_{\rm mat}\,\hat{D}_a\phi^a = \mathbb{S}^{ab} n_a \Phi_b + \mathcal{O}(r^2) = \mathbb{E}_\mathtt{EM}\cos\theta -2\,\mathbb{E}_\mathtt{EM}\,\frac{F(R_0)}{ R_0}\, (3\cos^2\theta-1) \,r+ \mathcal{O}(r^2).
\end{equation}
As argued in the discussion surrounding equation~(\ref{NormalMatterCorrection}), this is the net vertical stress (force per unit area) exerted on the field inside the RQF, and thus represents the buoyant forces due to matter. Surprisingly, at least to the order calculated, it is a sum of spherical harmonics that integrates to {\it zero} over the surface of the sphere. Setting $F(R)=1$ in equation~(\ref{TnPhiRN}) yields the corresponding result for a small round-sphere RQF sitting a fixed distance $R_0$ from a point charge in flat spacetime: no net buoyant force.

In going from flat spacetime to a RN black hole the net buoyant force remains zero, but there is an interesting temporal effect arising from the lapse function in the integrand on the left-hand side of equation~(\ref{quasilocal_Archimedes}). We find, in accordance with equation~(\ref{LapseSeries}), that
\begin{equation} \label{LapseRN}
N = 1 + \frac{A_0\cos\theta}{c^2} r  + \mathcal{O} (r^2),\;\;{\rm where}\;\;A_0=c^2F^\prime (R_0)=\frac{1}{F(R_0)}\left(\frac{GM}{R_0^2}-\frac{GQ^2}{c^2 R_0^3}\right).
\end{equation}
Here, $A_0$ is the acceleration of a fiducial point at the center of the sphere. The first term in parentheses is the corresponding Newtonian acceleration, and the second is a post-Newtonian correction that represents the well-known repulsive effect of the electric charge of a RN black hole. This lapse function says that, due to the acceleration of the RQF (or equivalently, the spacetime curvature), more proper time elapses at points in the top half of the sphere (furthest from the black hole), in a given parameter time interval, $dt$, than corresponding points in the bottom half. So although the magnitude of the upward buoyant forces in the top half is equal to the magnitude of the corresponding downward buoyant forces in the bottom half (at lowest order in $r$), there is a net upward buoyant {\it impulse}. Mathematically, the $\ell=1$ part of $N$ combines with the $\ell=1$ part of equation~(\ref{TnPhiRN}) to yield an $\ell=0$ term that does {\it not} vanish upon integration. We find
\begin{equation}
\int\limits_{\mathcal{S}_t} N \, d\hat{\mathcal{S}} \left[ \mathbb{S}^{ab} n_a \phi_b - {\rm P}_{\rm mat}\,\hat{D}_a\phi^a\right] = \int\limits_{\mathcal{S}_t} N \, d\hat{\mathcal{S}} \left[ \mathbb{S}^{ab} n_a \Phi_b + \mathcal{O}(r^2) \right]= \frac{\mathbb{E}_\mathtt{EM}}{c^2}\left(\frac{4\pi}{3} r^3\right) A_0+ \mathcal{O} (r^4),\label{LHS_quasilocal_Archimedes}
\end{equation}
which is the relativistic mass of the electrostatic field inside the RQF (to order $r^3$) times the mean acceleration of the frame, i.e., just equal in magnitude to the {\it weight} of the field inside, as we might expect. Integrating this {\it effective} net vertical buoyant force over parameter time $t$, as directed in equation~(\ref{quasilocal_Archimedes}), results in a nonzero net vertical buoyant {\it impulse} acting on the electrostatic field inside the RQF and supporting its weight.

The weight is calculated using the right-hand side of equation~(\ref{quasilocal_Archimedes}). In our example the twist is obviously zero ($\nu=0$), so we are left with two terms: $\frac{1}{c^2}\mathcal{E}\,\alpha^a\phi_a$ and ${\rm P}_{\rm geom}\,\hat{D}_a\phi^a$. The former is similar to the weight density in the local approach, $\frac{1}{c^2}\mathbb{E}\, a^a \Phi_a$, except that it contains vacuum contributions. We argued in the previous subsection that, in part, the geometrical pressure term provides a `geometrical buoyant force' that supports (cancels) these vacuum weight contributions. Let us quickly verify, in the present example, the facts on which this argument hinged. Beginning with equation~(\ref{GeometricalPressure}), a Maple calculation reveals that $\mathcal{P}^2 = 0$ (of course) and $\tilde{\mathcal{S}}^2 = \mathcal{O}(r^2)$, so these terms can be neglected relative to $^\mathcal{B}\mathcal{R}$ and $\mathcal{E}^2$, which both start at order $1/r^2$. Similarly, regarding the boundary Ricci scalar in equation (\ref{RicciScalar}), we verify that we can neglect the last two terms on the right-hand side since $\alpha_a \alpha^a = \mathcal{O}(1)$ (and $\nu = 0$), while $\hat{R} = 2/r^2$ (exactly) and $\hat{D}_a  \alpha^a$ starts at order $1/r$. With these details confirmed, we arrive at equation~(\ref{VacuumBuoyancyActual}), which verifies that the sum of the left-hand sides of equations~(\ref{VacuumWeight}) and (\ref{VacuumBuoyancy}) vanishes at orders $1/r^2$ and $1/r$, modulo spherical harmonic terms that integrate to zero. Using Maple we take the analysis two orders in $r$ higher, up to and including the order at which matter begins to contribute. The expressions contain a large number of spherical harmonic terms, but upon integration we arrive at a simple result for the (effective\footnote{The term ``effective" reminds us that the lapse function plays an important role here, i.e., properly speaking we are dealing with an impulse rather than a force.}) weight of the electrostatic field inside the RQF:
\begin{equation}\label{RHS_quasilocal_Archimedes}
\int\limits_{\mathcal{S}_t} N \, d\hat{\mathcal{S}} \left[\frac{1}{c^2}\mathcal{E}\,\alpha^a\phi_a  +{\rm P}_{\rm geom}\,\hat{D}_a\phi^a \right]= \frac{\mathbb{E}_\mathtt{EM}}{c^2}\left(\frac{4\pi}{3} r^3\right) A_0+ \mathcal{O} (r^4),
\end{equation}
in agreement with the effective buoyant force in equation~(\ref{LHS_quasilocal_Archimedes}).

\section{Summary and Conclusions}\label{Summary}

The local approach to constructing integrated energy and momentum conservation laws begins with the differential identity given in equation~(\ref{differential_local_conservation_law}), which reduces to $\nabla_a ( T^{ab}\Psi_b ) = T^{ab} \nabla_{(a} \Psi_{b)}$ in general relativity. The problems with this approach are essentially two-fold: (1) being local, this approach cannot account for gravitational effects, which are nonlocal (this is reflected in the fact that the identity is empty when $T^{ab}=0$), and (2) we get the intuitive form of a conservation law (namely, that the change in a physical quantity over time equals the net corresponding flux through the system boundary during that time) only when $\Psi^a$ is a Killing vector field, i.e., the approach relies on the existence of spacetime symmetries, which rules out spacetimes with interesting matter or gravitational dynamics.

A natural solution to the first problem is to replace the local matter stress-energy-momentum tensor, $T^{ab}$, with the Brown and York quasilocal stress-energy-momentum tensor, $T_\mathcal{B}^{ab}$, which represents both matter {\it and} gravitational fields, and which satisfies the differential identity $D_a (T_\mathcal{B}^{ab}\psi_b )=(D_a T_\mathcal{B}^{ab})\psi_b + T_\mathcal{B}^{ab}D_{(a}\psi_{b)}$ [equation~(\ref{differential_quasilocal_conservation_law})] in the boundary, $\mathcal{B}$~\cite{BY1993}. The solution to the second problem is to replace the notion of a rigid local frame (which, we argued, is required to make sense of an integrated conservation law, but does not always exist) with a rigid quasilocal frame (RQF) (which always exists)~\cite{EMM2009,EMM2011}. If $\psi^a$ is chosen to be $\frac{1}{c}u^a$ (where $u^a$ is the RQF observers' four-velocity) we get a completely general RQF {\it energy} conservation law for matter and gravitational fields in a finite system in general relativity (discussed in our previous paper~\cite{EMM2012}).  As shown in the present paper, if $\psi^a$ is chosen to be $-\frac{1}{c}\phi^a$, where $\phi^a$ is a stationary conformal Killing vector (CKV) field orthogonal to $u^a$ (the full set of six of which always exists) we get the completely general RQF {\it momentum} conservation law given in equation~(\ref{reduced_integrated_quasilocal_conservation_law}):
\begin{equation}
\int\limits_{\mathcal{S}_f - \mathcal{S}_i}  d\hat{\mathcal{S}} \, \left( \mathcal{P}^a+\frac{1}{c^2}\mathcal{S}^{ab}v_b\right)\phi_a =
\int\limits_{\Delta\mathcal{B}} N \, dt \, d\hat{\mathcal{S}} \left\{ \mathbb{S}^{ab} n_a \phi_b
- \left[\frac{1}{c^2}\mathcal{E}\,\alpha^a\phi_a +2\nu\epsilon_{ab}\mathcal{P}^a\phi^b + {\rm P}\,\hat{D}_a\phi^a \right]\right\},
\end{equation}
which is a conservation law for  linear or angular momentum according to whether $\phi^a$ is one of the three boost or three rotation CKV fields. The left-hand side of this equation gives the change in the momentum contained in the system as measured by the  RQF observers between the initial and final times.  The right-hand side is the flux of momentum  (including the gravitational momentum flux,  $-\frac{1}{c^2}\mathcal{E}\,\alpha^a\phi_a$) entering the system from the outside during this time. In short, we have constructed completely general energy~\cite{EMM2012} and momentum conservation laws that {\it do not rely on any spacetime symmetries}. The notion of an RQF plays a crucial role in the construction of these laws.

These new energy and momentum conservation laws teach us some new physics. They allow us to identify simple, exact, operational definitions for fluxes of gravitational energy~\cite{EMM2012} and momentum (linear and angular), and these fluxes in turn provide a deeper insight into what's really happening in a wide variety of physical phenomena. For instance, some simple, everyday effects in Newtonian mechanics are actually very tiny, subtle general relativistic effects multiplied by a large number (like $c^2/G$ or $c^4/G$) to become effects we see in the everyday world. When we drop an apple, for example, the apple gains energy and momentum relative to our accelerated frame of reference at rest on the Earth. The gain in energy is due to a gravitational energy flux involving the general relativistic effect of frame dragging~\cite{EMM2012}; the gain in momentum is due to a gravitational momentum flux involving the general relativistic effect that mass-energy has on the extrinsic curvature of the boundary of the reference frame.

Insofar as energy and momentum have traditionally been useful concepts in physics, we hope that these new, completely general RQF energy and momentum conservation laws will further deepen our understanding of energy and momentum, and find useful applications. Towards this end, we derived a general relativistic version of Archimedes' law, which we applied to understand how the weight of a small volume of electrostatic field, located a fixed distance from a Reissner-Nordstr\"{o}m black hole, is supported by the Maxwell stress buoyant forces of the surrounding electrostatic field. To our surprise, we found that the net buoyant force is actually zero, but the net buoyant {\it impulse} is not. The nonzero buoyant impulse is due to an acceleration-induced inhomogeneous time dilation at the boundary of the reference frame, as revealed by the RQF momentum conservation law.

\section*{Acknowledgments}

This work was supported in part by the Natural Sciences and Engineering Research Council of Canada.

\end{document}